\documentclass[numberedappendix, iop, apj]{emulateapj}
\usepackage{apjfonts}
\usepackage{natbib}
\usepackage{appendix}

\def\rhodotstar{$\dot{\rho_{*}}$}
\def\thetkern{$\theta_{\rm kern}$}
\def\arcsec{\hbox{$^{\prime\prime}$}}
\def\lya{Ly$\alpha$~}
\def\micron{$\mu$m}
\def\ciisl{\ion{C}{2}$^{*}$$\lambda$1335.7}
\def\rhodot{$\dot{\rho_{*}}$}
\def\smpykpc{$M_{\odot}{\rm \ yr^{-1} \ kpc^{-2}}$}

\def\aap{A $\&$ A}
\def\apjl{ApJL}
\def\apjs{ApJS}

\slugcomment{Accepted to ApJ, May 2 2011; published June 2011} 

\shorttitle{Star Formation in LBG outskirts}
\shortauthors{Rafelski, Wolfe, \& Chen}
 
\begin{document}

\title{Star Formation from DLA Gas in the Outskirts of Lyman Break Galaxies at z$\sim$3}

\author{Marc Rafelski\altaffilmark{1}, Arthur M. Wolfe\altaffilmark{1}, and Hsiao-Wen Chen\altaffilmark{2}}
\email{marcar@ucsd.edu} 
\altaffiltext{1}{Department of Physics and Center for Astrophysics and Space Sciences, UCSD, La Jolla, CA 92093, USA}
\altaffiltext{2}{Department of Astronomy \& Astrophysics, and Kavli Institute for Cosmological Physics, University of Chicago, Chicago, IL 60637, USA}

\begin{abstract}

We present evidence for spatially extended low surface brightness emission around Lyman break galaxies (LBGs) in the $V$ band image 
of the Hubble Ultra Deep Field, corresponding to the $z\sim3$ rest-frame FUV light, which is a sensitive measure of star formation rates 
(SFRs). We find that the covering fraction of molecular gas at $z\sim3$ is not adequate to explain the emission in the outskirts of LBGs, 
while the covering fraction of neutral atomic-dominated hydrogen gas at high redshift is sufficient. We develop a theoretical framework to 
connect this emission around LBGs to the expected emission from neutral \ion{H}{1} gas i.e., damped Ly$\alpha$ systems (DLAs), using the 
Kennicutt--Schmidt (KS) relation. Working under the hypothesis that the observed FUV emission in the outskirts of LBGs is from in situ 
star formation in atomic-dominated hydrogen gas, the results suggest that the SFR efficiency in such gas at $z\sim3$ is between factors 
of 10 and 50 lower than predictions based on the local KS relation. The total SFR density in atomic-dominated gas at $z\sim3$ 
is constrained to be $\sim10\%$ of that observed from the inner regions of LBGs. In addition, the metals produced by in situ star formation in 
the outskirts of LBGs yield metallicities comparable to those of DLAs, which is a possible solution to the ``Missing Metals'' problem for DLAs. 
Finally, the atomic-dominated gas in the outskirts of galaxies at both high and low redshifts has similar reduced SFR efficiencies and is 
consistent with the same power law. 
\end{abstract}

\keywords{ 
cosmology: observations ---
galaxies: evolution ---
galaxies: high-redshift ---
galaxies: photometry ---
general: galaxies ---
quasars: absorption lines}

\section{Introduction}

Understanding how stars form from gas is vital to our comprehension of galaxy formation and evolution. 
Although the physics involved in this process is not fully understood, 
we do know something about the principal sites where star formation occurs and where the gas resides. 
Most of the known star formation at high redshift occurs in Lyman break galaxies (LBGs), 
a population of star-forming galaxies selected for their opacity at the Lyman limit
and the presence of 
upper main sequence stars that emit FUV radiation. They also have very high star formation rates (SFRs) of
$\sim$80 $M_\odot$ yr$^{-1}$ 
after correcting for extinction \citep{Shapley:2003p4902}. 

While LBGs have a wide range of morphologies, the average half-light radius for $z\sim3$ LBGs  is 
about $\sim 2-3$ kpc in the optical at $\sim25$ mag \citep[e.g.,][]{Giavalisco:1996p4752, Law:2007p5043}.
However, studies of the Hubble Ultra Deep Field (UDF) have shown
that fainter LBGs have smaller half-light radii, around $\sim 1$ kpc for LBGs with brightnesses similar to 
those used in this study \citep[$V \sim 26-27$ mag;][]{Bouwens:2004p14601}. While we have no empirical knowledge 
about star formation in the outer regions of LBGs, simulations suggest that stars may be forming further out \citep[e.g.,][]{Gnedin:2010p12943}.
It still remains an unanswered question whether star formation 
occurs in the outer disks of high redshift galaxies. 
Local galaxies at $z\sim0$ are forming stars in their outer disks, as observed in the ultraviolet \citep{Thilker:2005p17091, Bigiel:2010b, Bigiel:2010a}.
At low redshift, this star formation occurs in atomic-dominated hydrogen gas \citep{Fumagalli:2008p5770, Bigiel:2010b, Bigiel:2010a},
where the majority of the hydrogen gas is atomic but molecules are present. 
At high redshift such gas resides in  damped Ly${\alpha}$ systems (DLAs).

DLAs are a population of \ion{H}{1} layers selected for their neutral hydrogen column 
densities of $N_{\rm H I} \geq2\times10^{20}$cm$^{-2}$, which dominate the neutral-gas content of the universe in the
redshift interval $0<z<5$. In fact, DLAs at $z\sim3$ contain enough gas to account for 25\%--50\% of the 
mass content of visible matter in modern galaxies \citep[see][for a review]{Wolfe:2005p382} and are
 neutral-gas reservoirs for star formation. 

The locally established Kennicutt--Schmidt (KS) relation \citep{Kennicutt:1998p3174,Schmidt:1959p12696} 
relates the SFR per unit area and the total gas surface density (atomic and molecular), 
$\Sigma_{\rm SFR}$ $\propto$ $\Sigma_{\rm gas}^{1.4}$.
While it is reasonable to use this relationship at low redshift in normal star-forming galaxies, 
many cosmological simulations use it at all redshifts
without distinguishing between atomic and molecular gas 
\citep[e.g.,][]{Nagamine:2004p5490, Nagamine:2007p462, Nagamine:2010p20920, Razoumov:2006p13906, Brooks:2007p16604, Brooks:2009p16590,
Pontzen:2008p5426, Razoumov:2008p13905, Razoumov:2009p11323, 
Tescari:2009p9594, Dekel:2009p6868, Dekel:2009p11011,  Keres:2009p11147,  
Barnes:2010p13908}\footnote{We note that there are also a large number of papers that do not assume the KS relation in their simulations and models
\citep[e.g.,][]{Kravtsov:2003p11341, Krumholz:2008p13733, Krumholz:2009p13722, Krumholz:2009p9796, Tassis:2008p19036,
Robertson:2008p19266, Gnedin:2010p12943, Gnedin:2011p27997, Feldmann:2011p27929}.}.
Yet at $z\sim3$, excluding regions immediately surrounding high surface brightness LBGs, 
the SFR per unit comoving volume, {\rhodot}, of DLAs was found to be less than 5$\%$ of 
what is expected from the KS relation \citep{Wolfe:2006p474}. 
This means that a lower level of in situ star formation occurs in atomic-dominated hydrogen gas
at $z\sim3$ than in modern galaxies\footnote{We note that the SFR efficiency discussed in this paper relates to the
normalization of the KS relation, and is not the same as the star formation efficiency (SFE), which 
is the inverse of the gas depletion time \citep[e.g.,][]{Leroy:2008p10197}.}. 
 
These results have multiple implications affecting such gas at high redshift. 
First, the lower SFR efficiencies in DLAs are inconsistent with the $158\mu$m cooling rates of 
DLAs with purely in situ star formation. Specifically, \citet{Wolfe:2008p5160} adopt the model of \citet{Wolfe:2003p2464},
in which star formation generates FUV radiation that heats the gas by the grain photoelectric mechanism.
Assuming thermal balance, they equate the heating rates to the [\ion{C}{2}] 158 {\micron} cooling rates of DLAs inferred
from the measured {\ciisl} absorption of DLAs \citep{Wolfe:2003p2464}. The DLA cooling rates exhibit a  bimodal 
distribution \citep{Wolfe:2008p5160}, and the population of DLAs with high cooling rates have inferred
heating rates significantly higher than that implied by the upper limits of FUV emission of spatially extended sources
 \citep{Wolfe:2006p474, Wolfe:2008p5160}. Therefore, another source of heat input is required, 
 such as compact star-forming regions embedded in the neutral gas;  e.g., LBGs. 
 Second, since {\rhodot} is directly proportional to the metal production rate, the limits on {\rhodot}
 shift the problem of metal overproduction in DLAs by a factor of 10 
 \citep[][; known as the ``Missing Metals'' problem for DLAs]{Pettini:1999p14867, pett04, Pettini:2006p11353, Wolfe:2003p2464},
to one of underproduction by a factor of three \citep{Wolfe:2006p474}. 
Lastly, the multi-component velocity structure of the DLA gas \citep[e.g.,][]{Prochaska:1997p11937} suggests that energy input by supernova
explosions is required to replenish the turbulent kinetic energy lost through cloud collisions; i.e.,
some in situ star formation should be present in DLAs. 

One possible way to reconcile the lack of detected in situ star formation in DLAs at high redshift and the properties of DLAs that require 
heating of the gas, is that compact LBG cores embedded in spatially extended DLA gas
may cause both the heat input, chemical enrichment, and turbulent kinetic energy observed in DLAs. 
There are several independent lines of evidence linking DLAs and LBGs; e.g. 
(1) there is a significant cross correlation between LBGs and DLAs \citep{Cooke:2006p467}, 
(2) the identification of a number of high-$z$ DLAs associated with LBGs 
\citep{Mller:2002p14362, Mller:2002p5373, Chen:2009p14317, Fynbo:2010p27721, Fumagalli:2010p19651, Cooke:2010p19462}, 
(3) the occurrence of Ly-$\alpha$ emission observed in the center of 
DLA troughs \citep{Mller:2004p13986, Cooke:2010p19462},
and (4) the appearance of DLAs in the spectra of rare lensed LBGs at high redshift, where the DLA and LBG have similar redshifts \citep{Pettini:2002p11290, Cabanac:2008p19014, DessaugesZavadsky:2010p18969}.

In fact, recent results are consistent with the idea that gas in spatially extended DLAs
encompasses compact LBGs. 
\citet{Erb:2008p2967} demonstrated that LBGs are rapidly
running out of ``fuel'' for star formation, and cold ($T\sim$100 K) gas in DLAs is a natural fuel source. 
This finding is supported by the measurements of the SFR and gas densities of LBGs by \citet{Tacconi:2010p12798}. 
We note that if DLAs are the fuel source for the LBGs, 
the DLAs in turn would have to be replenished since the comoving density of DLAs at its peak ($z\sim3.5$)
is about 1/3 the current cosmic mass density of stars \citep{Prochaska:2009p11257}.
Presumably, they are replenished through accretion of warm ($T\sim10^4$K) ionized flows\footnote{Often referred to as ``cold'' flows, 
but we call them warm flows since cold refers to $T\sim$100K gas in this paper.}
 \citep{Dekel:2006p14369, Dekel:2008p14374, Dekel:2009p6868, Dekel:2009p11011, Bauermeister:2010p13833}.
 
While LBGs embedded in spatially extended DLA gas help resolve some properties of DLAs, 
it is problematic whether metal-enriched outflows from LBGs can supply the 
required metals seen in DLAs. Nor is it clear whether such outflows can generate turbulent kinetic energy at rates
sufficient to balance dissipative losses arising from cloud collisions implied by the multi-component
velocity structure of DLAs\footnote{The cloud crossing time is $t_{\rm cross}=H/v$ and the cloud collision time 
is $t_{\rm coll} \simeq \frac{1}{n\sigma v}$, where $H$ is the DLA scale height, $v$ is the cloud velocity, $n$ is the number density of clouds,
and $\sigma$ is the geometric cross section of the clouds. Therefore, the ratio of the cloud crossing time to the cloud collision time is 
$\frac{t_{\rm cross}}{t_{\rm col}}\simeq n\sigma H \sim \tau$, where $\tau$ is the optical depth. Consequently, if
$\tau>1$ as is observed, the clouds would dissipate on a timescale short compared to the crossing time \citep[e.g.,][]{McDonald:1999p19027}. }.
We note that \citet{Fumagalli:2011p22356} show that
that the filamentary gas structures in the cold mode accretion scenario that provide galaxies with fresh fuel 
\citep[e.g.,][]{Dekel:2009p6868, Keres:2009p11147}
are not sufficiently dense to produce DLA absorption, nor do these filaments have a large enough
area covering fraction \citep{FaucherGiguere:2011p20792}. The only location with sufficient covering fraction and
high enough densities for the gas to become self shielded is in the vicinity around galaxies (2-10kpc).

To address these issues, we adopt the working
hypothesis that  in situ star formation occurs in
the presence of atomic-dominated gas in the outskirts of LBGs,
similar to the outer disks of local galaxies. 
Since most of the atomic-dominated gas at high redshift is in DLAs,
we assume that this is DLA gas. 
We emphasize that while the past results \citep{Wolfe:2006p474} set sensitive upper limits on  in situ  star formation
in DLAs without compact star-forming regions like LBGs, no such limits exist for DLAs containing such objects.  

For these reasons we search for spatially extended star formation associated with LBGs at  $z\sim3$
by looking for regions of low surface brightness (LSB) emission surrounding the LBG cores. 
We are searching for in situ star formation on scales up to $\sim10$ kpc, where the detection of faint emission would indicate the presence
of spatially extended star formation. 
It would also uncover a mode of star formation hitherto unknown at high $z$, 
and could help solve the dilemmas cited above. 
We test the hypotheses that 
(1) the extended star formation is fueled by atomic-dominated gas as probed by the DLAs, and 
(2) star formation occurs at the KS rate.
That is, we consider whether star formation occurs in the outskirts of LBGs, whether that star formation occurs
in atomic-dominated gas, and at what SFR efficiency the stars form.  

This paper is organized as follows. In Section 2, we describe the observations used, and in Section 3 we 
we identify a sample of compact LBGs at $z\sim3$ to search for extended LSB 
emission around the compact cores of the LBGs. In Section 4, 
we describe the image stacking technique, measure a median radial profile of the extended LSB emission, 
and discuss possible selection biases of the observations.
In Section 5, based on the observed radial profile, 
we calculate the corresponding SFR surface density distribution and the sky covering fraction of the extended LSB emission.  
We then calculate the in situ SFR density and metal production in these extended LSB regions.
 In Section 6, we develop a theoretical framework to connect known DLA statistics to the observed surface density distribution of SFR 
 in the outskirts of LBGs, and to obtain an empirical estimate of the star formation efficiency in distant galaxies.
 In Section 7, we discuss the impacts of our analysis in understanding the star formation relation in the distant universe, 
 and in interpreting the observed low metal content of the DLA population.
 
Throughout this paper, we adopt the AB magnitude system and an 
$(\Omega_M, \Omega_\Lambda, h)=(0.3,0.7,0.7)$ cosmology.


\begin{figure*}[t!]
\center{
\includegraphics[scale=0.64, viewport=15 5 490 350,clip]{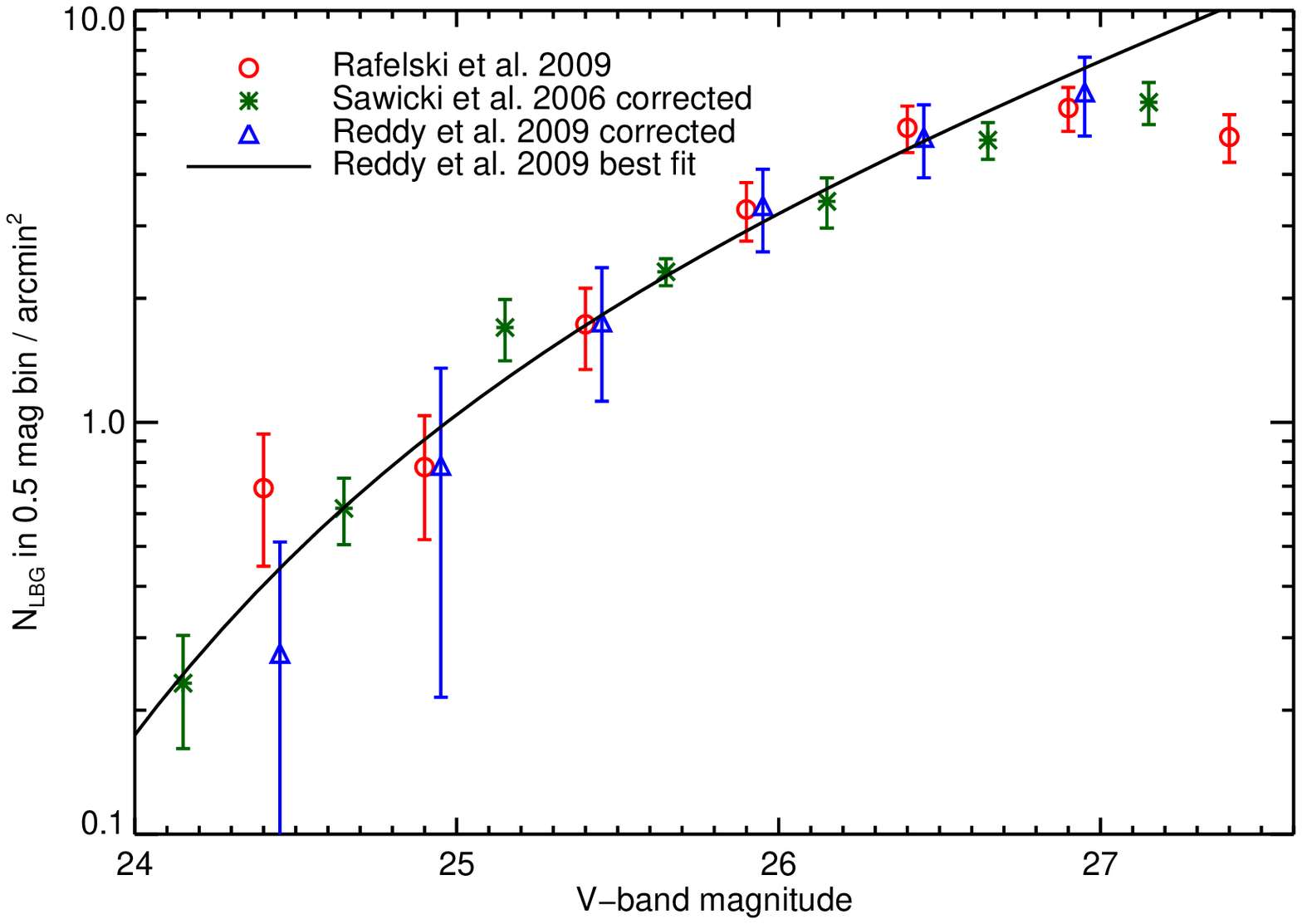}}
\caption{Comparison of the number counts of $z\sim3$ LBGs in 0.5 mag bins per square arcminute.
The red circles are from \citet{Rafelski:2009}, the green crosses are from \citet{Sawicki:2006p1733}
 corrected for completeness, and the blue triangles are from \citet{Reddy:2009p6997}
also corrected for completeness. 
The points from \citet{Reddy:2009p6997} are offset by 0.05 mag in order to avoid
overlapping the \citet{Rafelski:2009} points. 
The black line is the best-fit Schechter function \citep{Schechter:1976p12758}
from \citet{Reddy:2009p6997} with $\alpha=-1.73\pm0.13$, $M^*_{\rm AB}(1700\AA)=-20.97\pm0.14$, 
and $\phi^*=(1.71\pm0.53$)$\times10^{-3}$ Mpc$^{-3}$.
}
   \label{fig:numcts}
\end{figure*}

\section{Observations}

We implement our search for spatially extended star formation around $z\sim3$ LBGs in the most sensitive high-resolution images available: 
the $V$ band image of the UDF taken with the {\it Hubble Space Telescope (HST)}. 
This image is ideal because (1) of its high angular resolution (PSF FWHM = 0.{\arcsec}09), (2) at $z\sim3$ the $V$ band fluxes correspond
to rest-frame FUV fluxes, which are sensitive measures of SFRs, 
since short-lived massive stars produce the observed UV photons,
and (3) the 1$\sigma$ point source limit of $V$=30.5 implies high sensitivity.
Since most of the LBGs in the UDF are too faint for spectroscopic identification, the $u$ band is  needed to find $z\sim3$ LBGs via their flux 
decrement due to the Lyman limit through color selection and photometric redshifts. To this end, we acquired one of the most sensitive 
$u$ band images ever obtained and identified 407 LBGs at $z\sim3$ \citep[][see also \citet{Nonino:2009p9926}]{Rafelski:2009}.

Throughout the paper we utilize the $B$, $V$, $i^\prime$, and $z^\prime$ band 
(F435W, F606W, F775W, and F850LP, respectively) observations of the  UDF
\citep{Beckwith:2006p1529}, obtained with the Wide Field Camera 
on the {\it HST} Advanced Camera for surveys \citep[ACS;][]{Ford:2002p6197}. These images cover 12.80 arcmin$^{2}$, 
although we only use the central 11.56 arcmin$^{2}$ which overlaps the $u$ band image
from \citep{Rafelski:2009}. The $u$ band image was  obtained with the Keck I telescope and the 
blue channel of the Low-Resolution Imaging Spectrometer 
\citep[LRIS;][]{Oke:1995p6046,McCarthy:1998p6102} and has a 1$\sigma$ depth of 30.7 
mag arcsec$^{-2}$ and a limiting magnitude of 27.6 mag. The sample described below also 
makes use of the observations taken with the NICMOS camera NIC3 in the $J$ and $H$ bands 
 \citep[F110W and F160W;][]{Thompson:2005p4915} whenever the field of view (FOV) overlaps.

\section{Sample Selection}

In order to search for spatially extended star formation associated with $z\sim3$ LBGs, we require 
a sample of such galaxies to form a super-stack of LBG images that we describe here. In Section 3.1, 
we compare the number counts of the $z\sim3$ LBGs in the UDF to those in the literature.
Then in Section 3.2, we select a subsample that is appropriate for stacking in order to improve the signal-to-noise (S/N) 
in the LBG outskirts as described in Section 4. Lastly, in Section 3.3 we investigate possible selection biases of the observations.

The samples in this paper are based on the $z\sim3$ LBG sample of \citet{Rafelski:2009}, which 
contains 407 LBGs selected by using a combination of photometric redshifts and the $u$ band drop out technique
\citep{Steidel:1992p1911,Steidel:1995p1873, Steidel:1996p5981, Steidel:1996p5985}. 
This selection of LBGs is enabled by the extremely deep $u$ band image described in Section 2, needed
to reduce the traditional degeneracy of colors between $z\sim3$ and $z\sim0.2$ galaxies that can
yield incorrect redshifts at $z\sim3$ without the $u$ band 
\citep{Ellis:1997p3771, FernandezSoto:1999p2784, Benitez:2000p3572, Rafelski:2009}. 
\cite{Rafelski:2009} found that the resultant sample is likely to have
 a contamination fraction of only $\sim3\%$. Any such contamination will have a minimal effect on 
our results and is included in the uncertainties (see Section 4.3). In addition, \citet{Rafelski:2009} found that the 
LBG sample is complete to $V\sim27$ mag, limited by the depth of the $u$ band image.

 \begin{figure*}[t!]
\center{
\includegraphics[scale=0.42]{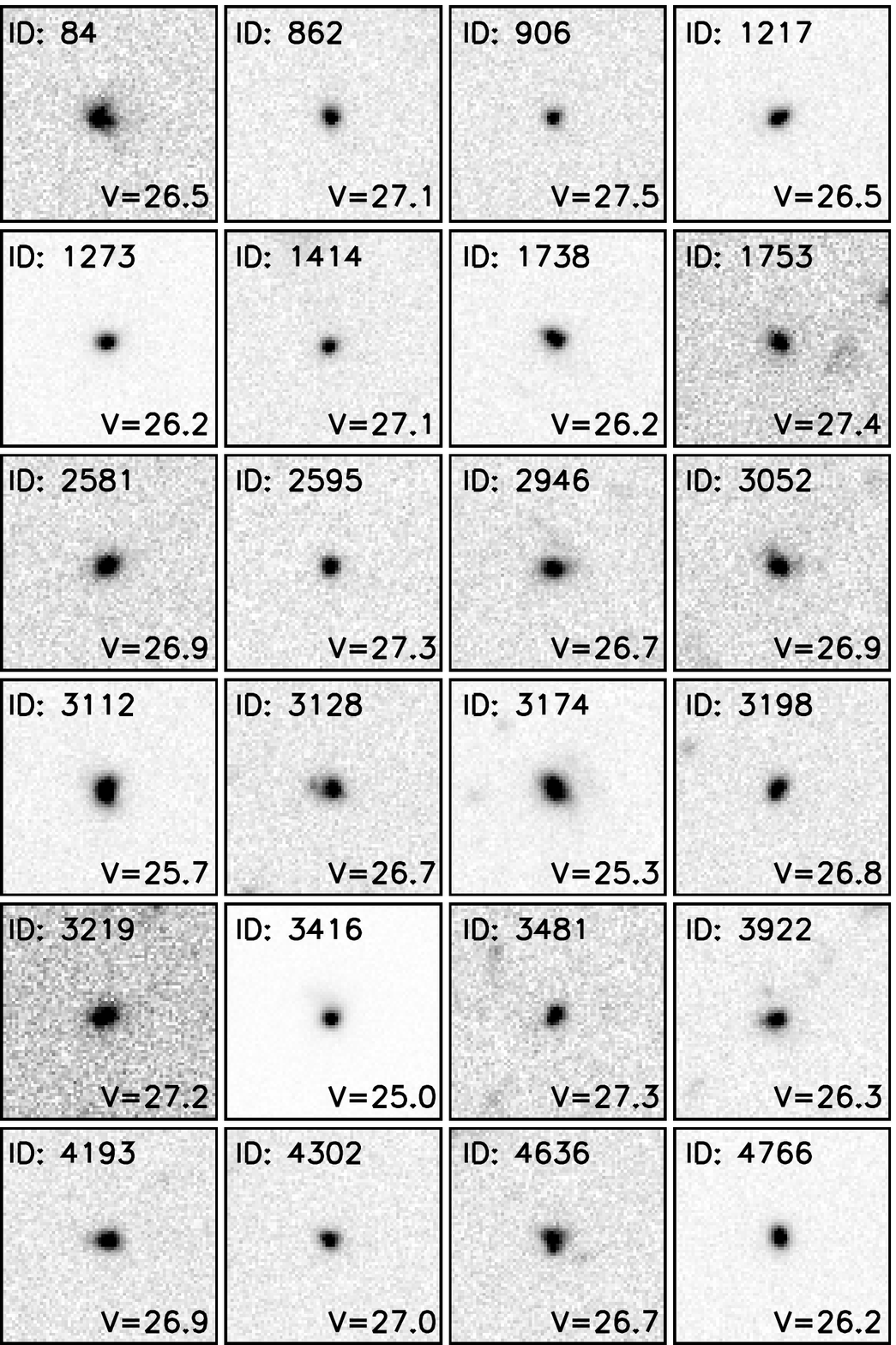}
\hspace{5mm}
\includegraphics[scale=0.42]{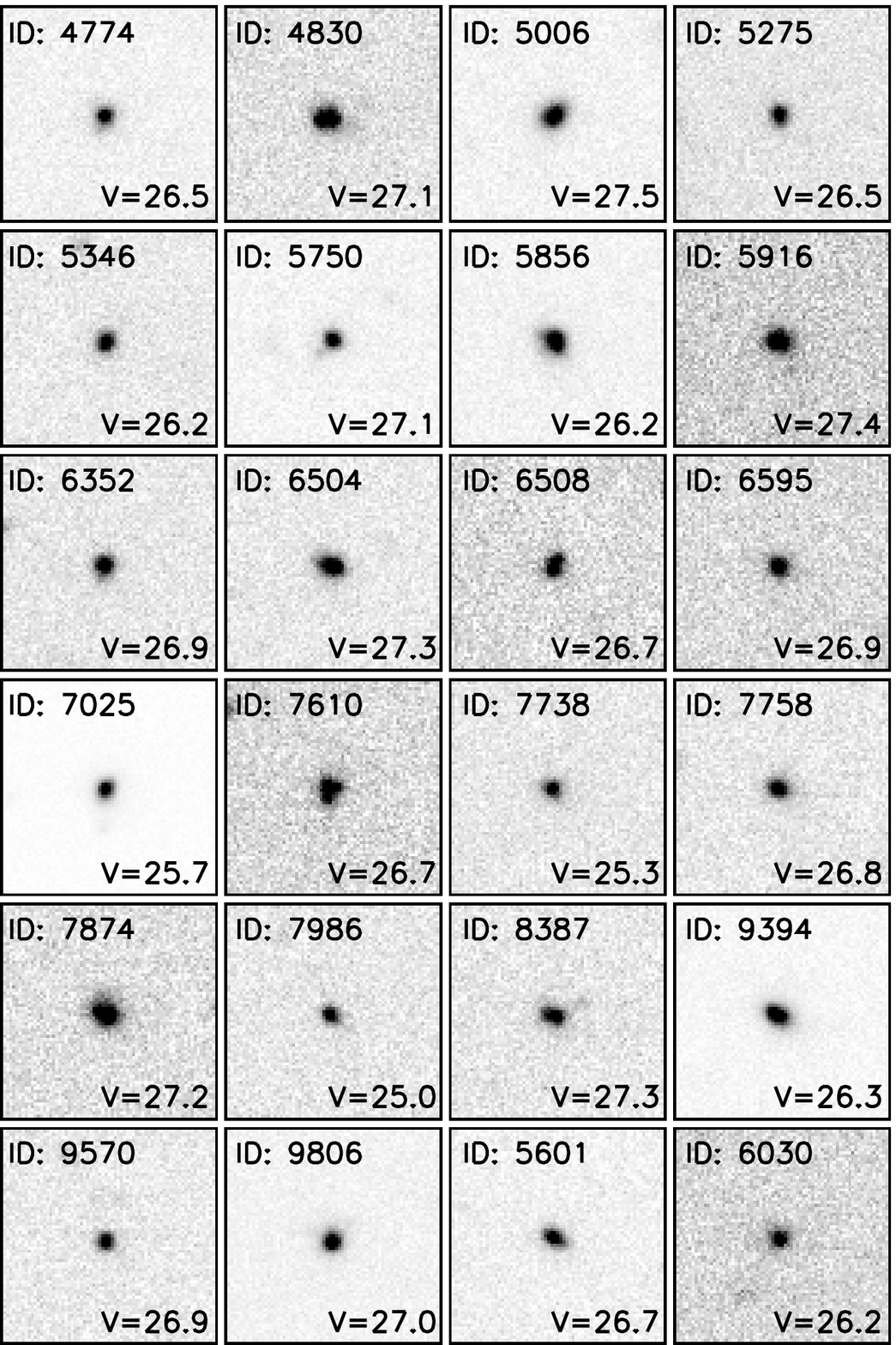}
}
\caption{Thumbnail images in the $V$ band of the 48 $z\sim3$ LBG subsample. 
The thumbnails are 2.4 arcsec on a side, the same size as the composite image in Figure \ref{fig:stack}, which corresponds to $18.5$ kpc at $z\sim3$.}
   \label{fig:thumb}
\end{figure*}

\subsection{Number Counts}

In order to (1) be confident in our sample selection, (2) verify that we are probing the correct comoving volume,
and (3) make completeness corrections in Sections 5 and 6 (as described in Appendix A), 
we compare the number counts of the $z\sim3$ LBG selection from \citet{Rafelski:2009}  to the completeness
corrected number counts from \citet{Sawicki:2006p1733} and \citet{Reddy:2009p6997} in Figure
\ref{fig:numcts}. The uncertainties shown from \citet{Rafelski:2009} are only poisson and therefore
have smaller error bars than those by \citet{Reddy:2009p6997}. 
The points from \citet{Reddy:2009p6997} are offset by 0.05 mag for clarity,
and their uncertainties include both poisson and field to field variations.
The number counts and the best-fit Schechter function \citep{Schechter:1976p12758} from
\citet{Reddy:2009p6997} are converted to number of LBGs per half-magnitude bin 
using a similar conversion as in \citet{Reddy:2008p4837}. Specifically,
we use the $R$ band as a tracer of rest-frame 1700$\AA$ emission.
The apparent measured $R$ magnitude, $R_{\rm AB}$, is converted from
the absolute magnitude using the relation:
\begin{equation}
R_{\rm AB} = M_{\rm AB}(1700{\rm \AA}) + 5 {\rm log}(d_L/ 10 {\rm pc}) +2.5{\rm log}(1+z)\;,
\label{eq:mag}
\end{equation}
where $M_{\rm AB}(1700\AA)$ is the absolute magnitude at the rest-frame 1700$\AA$,
and $d_L$ is the luminosity distance. 
We apply the $k$-correction from \citet{Rafelski:2009} of $R_{\rm AB}-V_{\rm AB}$= $-$0.15 mag to get the $V$ band mag. 
We use a comoving volume of  26436~Mpc$^3$ for the redshift interval 2.7$<z<$3.4 and 
an area of 11.56~arcmin$^2$ for the number count conversion. 
We adopt a Schechter function with parameters found by 
\citet{Reddy:2009p6997} of $\alpha=-1.73\pm0.13$, 
$M^*_{\rm AB}(1700\AA)=-20.97\pm0.14$, and $\phi^*=(1.71\pm0.53$)$\times10^{-3}$ Mpc$^{-3}$.

The resultant number counts agree nicely, and show that the completeness of the 
\citet{Rafelski:2009} $z\sim3$ LBG sample
matches the previous completeness limit found of $V\sim27$ magnitude. 
The agreement also suggests that the comoving volume
for the redshift interval 2.7$<z<$3.4 is appropriate for the sample, which is important for the argument given in Section 5. 
More importantly, the agreement of the number counts allows us to use the best-fit Schechter function 
from \citet{Reddy:2009p6997} to determine the expected number of LBGs at fainter magnitudes, 
providing the needed information to make completeness corrections in Appendix A. We note that this luminosity function
is valid to $R\sim26.5$, after which the extrapolation to fainter magnitudes could be a potential source of error, and we 
address this below. 

\begin{deluxetable*}{crrrrrcc}
\tabletypesize{\scriptsize}
\tablecaption{Properties of LBGs Included in Image Stack 
\label{tab1}}
\tablewidth{0pt}
\tablehead{
\colhead{ID\tablenotemark{a}} &
\colhead{$z_{\rm phot}$\tablenotemark{b}} &
\colhead{V} &
\colhead{$u-V$} &
\colhead{$B-V$} &
\colhead{$V-z^\prime$} &
\colhead{FWHM} &
\colhead{Ellipticity} \\
\colhead{} &
\colhead{} &
\colhead{(mag)} &
\colhead{(mag)} &
\colhead{(mag)} &
\colhead{(mag)} &
\colhead{(arcsec)} &
\colhead{}
}

\startdata

84 &3.11$_{-0.40}^{+0.40}$ &26.56$ \pm $0.02 &2.40$ \pm $0.35 &0.70$ \pm $0.05 &0.03$ \pm $0.04 &0.24 &0.18 \\
862 &3.18$_{-0.41}^{+0.41}$ &27.16$ \pm $0.02 &1.87$ \pm $0.36 &0.74$ \pm $0.05 &-0.44$ \pm $0.05 &0.12 &0.14 \\
906 &2.68$_{-0.36}^{+0.36}$ &27.52$ \pm $0.02 &1.50$ \pm $0.36 &-0.10$ \pm $0.04 &-0.14$ \pm $0.05 &0.10 &0.04 \\
1217 &2.75$_{-0.37}^{+0.37}$ &26.53$ \pm $0.01 &2.28$ \pm $0.30 &0.34$ \pm $0.02 &0.01$ \pm $0.02 &0.14 &0.23 \\
1273 &3.03$_{-0.39}^{+0.39}$ &26.24$ \pm $0.01 &2.79$ \pm $0.37 &0.58$ \pm $0.03 &0.02$ \pm $0.02 &0.12 &0.12 \\
1414 &2.86$_{-0.38}^{+0.38}$ &27.19$ \pm $0.02 &1.82$ \pm $0.37 &0.49$ \pm $0.06 &-0.03$ \pm $0.05 &0.10 &0.13 \\
1738 &2.67$_{-0.36}^{+0.36}$ &26.27$ \pm $0.01 &1.05$ \pm $0.08 &0.03$ \pm $0.02 &-0.30$ \pm $0.03 &0.16 &0.15 \\
1753 &3.44$_{-0.43}^{+0.43}$ &27.47$ \pm $0.03 &1.02$ \pm $0.35 &1.39$ \pm $0.14 &0.58$ \pm $0.04 &0.18 &0.12 \\
2581 &3.40$_{-0.43}^{+0.43}$ &26.92$ \pm $0.02 &2.05$ \pm $0.35 &1.04$ \pm $0.07 &0.43$ \pm $0.03 &0.18 &0.24 \\
2595 &2.97$_{-0.39}^{+0.39}$ &27.37$ \pm $0.02 &1.60$ \pm $0.35 &0.22$ \pm $0.04 &-0.17$ \pm $0.05 &0.13 &0.05
\enddata

\tablecomments{$V$ magnitudes are total AB magnitudes, and colors are isophotal colors. $u$ band photometry is from \citet{Rafelski:2009} and the rest are from \citet{Coe:2006p1519}.
Non-detections in the $u$ band are given 3$\sigma$ limiting magnitudes. 
This table is available in its entirety in a machine-readable form in the online journal. A portion is shown here for guidance regarding its form and content. }
\tablenotetext{a}{ID numbers from \citet{Rafelski:2009} which match those by \citet{Coe:2006p1519}.}
\tablenotetext{b}{Bayesian Photometric Redshift (BPZ)  and uncertainty from 95\% confidence interval from \citet{Rafelski:2009}.}

\end{deluxetable*}

\subsection{Catalogs of LBGs and Stars}

We use two catalogs of $z\sim3$ LBGs in this paper. 
The first is the full sample of 407 $z\sim3$ LBGs as described in \citet{Rafelski:2009}. The
sample redshift distribution is shown in Figure 12 of \citet{Rafelski:2009} and has a mean 
photometric redshift of $3.0\pm0.3$. 
The second (hereafter referred to as ``subset sample'') 
is a sample of $z\sim3$ LBGs selected to create a composite image to improve 
the signal-to-noise of the surface brightness profile described below. These LBGs 
are selected to be compact, symmetric, and isolated, similar to the selection done at higher redshift 
by \citet{Hathi:2008p4334}. The LBGs are selected to be compact and symmetric to aid in stacking
LBGs of similar morphology and physical characteristics such that the bright central regions of the 
LBGs overlap. They were also selected to be isolated from nearby neighbors to avoid 
coincidental object overlap and dynamically disturbed objects. 

To select objects that are compact, symmetric, and isolated, we measure morphological parameters
in the $V$ band image using $\tt SExtractor$ \citep{Bertin:1996p6133}. We experimented with different
morphological parameters, such as asymmetry \citep{Schade:1995p12525}, 
concentration \citep{Abraham:1994p9329, Abraham:1996p12567},
Gini coefficient \citep{Lotz:2004p5115}, and clumpiness \citep{Conselice:2003p5100}. 
However, we found that the best sample was selected based on the FWHM for compactness and ellipticity 
$\epsilon = (1-b/a)$ for symmetry, similar to the criteria in \citet{Hathi:2008p4334}.
Specifically, for the subset sample, we require that FWHM $\leq$ 0.{\arcsec}25 and $\epsilon \leq0.25$, a slightly
more conservative selection than \citet{Hathi:2008p4334}.
Lastly, to select isolated objects, we require that  there are no other objects within 1.{\arcsec}4 brighter than 29th mag. 
These requirements yield a sample of 48 LBGs, representing $\sim12\%$ of the $z\sim3$ LBG sample, whose 
properties are compared in Section 3.3.
We show this sample as thumbnails that are 2.4 arcsec on a side, which corresponds to 18.5 kpc at $z\sim3$, in Figure \ref{fig:thumb},
and a table with relevant information about the 48 LBGs in Table \ref{tab1}. Information on the full sample of 407 LBGs is available
in \citet{Rafelski:2009}. 

In addition to the two samples of LBGs, we also need a sample of stars for an accurate measurement
of the point spread function (PSF) of the UDF images. We obtain this star sample from
 \citet{Pirzkal:2005p9469}, using only those stars with confirmed grism spectra, which mostly consist of 
 M dwarfs. We exclude the stars that are saturated, leaving 15 stars in the $V$ band image within our FOV with 
 $V=25.7\pm1.3$, more than adequate to measure the PSF. 
 We note that a comparison of the star PSF with the LBGs shows that they are all resolved in the high resolution ACS images.

\subsection{Comparison of the Subset and Full LBG Samples}

We investigate whether the subset sample of 48 LBGs is drawn from the same parent population of the full sample of 407 LBGs
by comparing the magnitude, color, and redshifts of two the samples. 
First, we find little variation in the magnitude distributions of the two samples, with a difference in the mean of $\sim0.3$ mag. 
The subset sample is somewhat fainter, with the full sample having an average AB magnitude of $V=26.4\pm0.9$ and the subset sample with $V=26.7\pm0.6$.
That is, there is a minor systematic selection of fainter LBGs in the subset sample, although this difference
is not significant. The similar magnitude distribution of the rest-frame FUV flux suggests that the SFR of the 
two samples is similar. 

Second, we compare the mean colors of the of the two samples and find the two
samples have the same colors. We test this both for the distribution and for stacks of the LBGs. 
First, the mean of the distribution of the subset sample yields colors of 
$B-V$ = 0.5$\pm0.4$, $V-i^\prime$ = 0.0$\pm0.2$, and $i^\prime-z^\prime$ =0.0$\pm0.1$, 
while the full sample has colors of
$B-V$ = 0.6$\pm0.4$, $V-i^\prime$ = 0.1$\pm0.2$, and $i^\prime-z^\prime$ =0.0$\pm0.1$.
Second, the color of the stacked subset sample based on aperture photometry has colors of
$B-V$ = 0.2$\pm0.2$, $V-i^\prime$ = 0.1$\pm0.2$, and $i^\prime-z^\prime$ =0.1$\pm0.2$, 
while the full sample stack has colors of
$B-V$ = 0.3$\pm0.1$, $V-i^\prime$ = 0.1$\pm0.1$, and $i^\prime-z^\prime$ =0.2$\pm0.1$.
These colors are not significantly different based on both the distribution and the stacked photometry uncertainties.
The similar distribution of colors suggests that the two samples are made of the same stellar populations 
and that their star formation histories (SFHs) are similar. 

Lastly, the two samples have very similar redshift distributions, with the same mean redshift of $3.0\pm0.3$. 
We therefore conclude that the subset sample and full sample of LBGs are 
equivalent in magnitude, color, and redshift, and therefore are probably drawn from the same parent population of LBGs
that have similar SFRs, stellar populations, and SFHs. 
Hence, we are relatively confident that the results determined below for the subset sample of LBGs are 
applicable to the full sample. 

For the sake of completeness, we also consider stacking the full stack of LBGs
in Appendix B. We note that stacking the full sample introduces contamination into the stack, such as nearby galaxies.  
In addition, bright parts of morphologically different galaxies contribute to the faint parts of other galaxies. We therefore
do not use this as our primary stack, and take the full stack as an upper limit to the emission from the full sample of LBGs. 

\section{Analysis of UDF Images}

The $V$ band image of the UDF is the most sensitive high resolution image
covering the rest-frame FUV at $z\sim3$ available. 
However, even this image does not reach the desired sensitivity to search for
spatially extended star formation on scales up to $\sim$10 kpc (as shown below). We wish to 
increase the S/N high 
enough to probe down to low values of the SFR surface density ($\Sigma_{\rm SFR}$). 
Image stacking methods can be used to study the average properties
of well defined samples in which individual objects do not have the necessary S/N
\citep[e.g.,][]{Pascarelle:1996p12752, Zibetti:2004p12747, Zibetti:2005p886, Zibetti:2007p16154, Hathi:2008p4334}.

We therefore create super-stacks of the LBG images in Section 4.1 and investigate how the sky-subtraction uncertainty affects those stacks in Section 4.2.
Using the super-stack of the subset sample of LBGs described in Section 3.2, we determine the radial surface brightness profile of $z\sim3$ LBGs 
in Section 4.3, which will be used for much of the analysis throughout the paper. Lastly, in Section 4.4, we investigate the effects of the 
{\lya} line on the stacked image.

\subsection{Image Stacks}

\begin{figure}[t]
\center{
\includegraphics[scale=0.52]{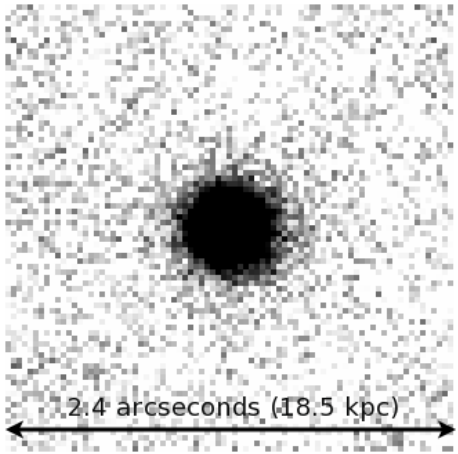}}
\caption{Composite image in the $V$ band of the 48 $z\sim3$ LBG subsample. 
Each stamp is 2.4 arcsec on a side, which corresponds to  $18.5$ kpc
at $z\sim3$.}
   \label{fig:stack}
\end{figure}

We create stacked composite images for the LBG subset sample, full LBG sample, and stars sample using custom IDL code.
For each object, we fit a two-dimensional Gaussian using $\tt MPFIT$ \citep{Markwardt:2009p27992} to determine
a robust center.  We then shift each object to be centered with sub-pixel resolution, interpolating with a damped sinc function. We
then create thumbnail images for each object and combine all the objects by taking the median of all the thumbnails, yielding a 
robust stacked image which is not sensitive to outliers. 
While some of the individual thumbnails may have faint emission regions below the level of the isolation criteria,
they do not contribute to the median because such emission would need to occur at the same pixels for a significant number
objects to affect the median. Therefore, independent of the origin of any such faint regions, they do not affect their median, and therefore
do not affect the final stack. 
While we are most interested in the $V$ band, we also carry out this procedure 
for the $B$, $V$, $i^\prime$, and $z^\prime$ bands and the stars sample.
Figure \ref{fig:stack} shows the stacked $V$ band image, as this corresponds to the rest-frame FUV 
luminosity at $z\sim3$ which is a sensitive measure of the SFR. The image is 2.4 arcsec on a side, which corresponds to 18.5 kpc at $z\sim3$.

We showed in Section 3.3 that the subset sample is representative of the full LBG sample and here investigate
any possible variations in the radial surface brightness profile with magnitude and FWHM within the subsample itself. 
We check if these parameters affect the stack by creating three independent stacks of 
different brightnesses or FWHM. In the case of magnitude, 
we create one stack of the brightest 16 LBGs, a second stack of the next 16 LBGs, and a third stack with the faintest 16 LBGs,
and repeat for FWHM.
We find the profiles to be very similar and find that the magnitude and FWHM range does not affect our stack. 
We also investigate the change in the surface brightness profile color. We find that the 
variations of the LBG composite images for the $B$, $V$, $i^\prime$, and $z^\prime$ bands
across radius are small compared to their uncertainties, and no clear change in color is obvious at any
radius. 

We also test the difference between
taking the median and the mean of the images in our stack. The profiles of the mean stack are slightly higher at the bright end, but are very similar to 
the median stack starting at $\sim$0.{\arcsec}3. We chose to go with the median as it is more robust to possible contamination in the outskirts. 
In this way, we are not sensitive to contamination if it only occurs in a small subset of our sample. 

In creating a stack of the star data to measure the PSF, 
we scale each star to the peak of the fitted two-dimensional Gaussian before stacking. 
For the PSF, we only care about the shape of the PSF, and not the actual value of the flux. 
Given the wide distribution of magnitudes of the stars,
this scaling improves the PSF determination. 
We do not scale the LBGs before stacking, 
as we care about the actual measured flux in the outskirts of the LBGs.
We find that scaling does not have a significant effect on the LBG 
stack due to the small range in brightnesses of the selected LBGs, with the 
radial surface brightness of the two stacks being consistent within the uncertainties,
and this would therefore not alter our results.

\subsection{Sky-subtraction Uncertainty}

We hope to accurately characterize the sky background and the uncertainties due to the subtraction of this sky background. 
The sky-subtraction uncertainty of the UDF was carefully investigated by \citet{Hathi:2008p4334}, and we follow their 
prescription for determining the $1\sigma$ sky-subtraction error. \citet{Hathi:2008p4334} found
that it was more reliable to characterize the sky locally rather than globally for the entire image. 
For this reason, we redetermine the local sky background for the 407 $z\sim3$ LBGs in our sample. 
First, we measure the sky background in each of the thumbnail images 
of the full LBG sample using IDL and the procedure 
$\tt MMM.pro$\footnote{Part of the IDL Astronomy User's Library}, adapted from the DAOPHOT routine 
by the same name \citep{Stetson:1987p12673}. This procedure iteratively determines the background, 
removing low probability outliers each time, until the sky background is determined. We then find
the $1\sigma$ sky value by fitting a Gaussian to the distribution of sky values. For the $V$ band, this
gives us a value for $\sigma_{\rm sky,ran}$ of $3.52\times10^{-5}$ electrons s$^{-1}$, very similar to that
found by \citet{Hathi:2008p4334} of $3.55\times10^{-5}$ electrons s$^{-1}$.  Using this number
and the formalism in \citet{Hathi:2008p4334}, we find a 1$\sigma$ sky-subtraction error of 
30.01 mag arcsec$^{-2}$. 

If the error is random, the uncertainty of the sky-subtraction will decrease as we stack more images together. 
In fact, for a median stack in the Poisson limit, the 1$\sigma$ uncertainty is $1.25 / \sqrt{N}$, where $N$ is the number of images. 
We test this relation specifically for the UDF data as the error may not be completely random. We stack
48 blank thumbnails and compare the standard deviation of stacked pixels to the median of the standard 
deviations of each individual thumbnail and find the above relation holds to 99\% accuracy. We are therefore
confident that the sky-subtraction noise decreases with stacking as expected, yielding larger S/N values.
For the stack of 48 LBGs from our subset sample, we find a 1$\sigma$ sky-subtraction error of 
31.87 mag arcsec$^{-2}$. We use both of these sky-subtraction errors as our sky limits below in Section 4.3.

\subsection{Radial Surface Brightness Profile}

\begin{figure}[t]
\center{
\includegraphics[scale=0.55, viewport=30 25 460 330,clip]{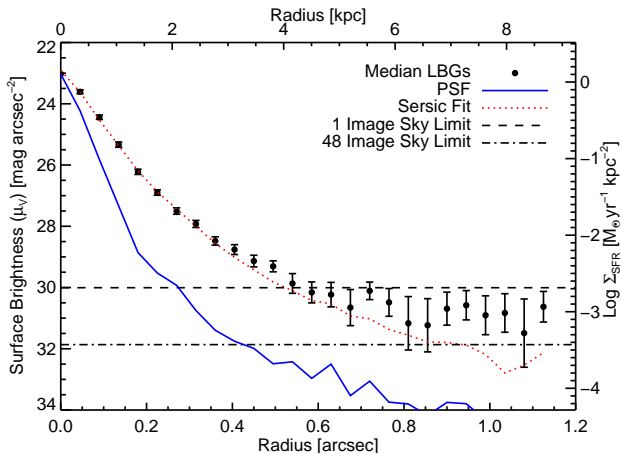}}
\caption{Extracted surface brightness profile from the stacked image, where the black points are the 48 LBG composite
profile and the blue line is the measured point-spread function from stars. The dotted red line is the best-fit S\'{e}rsic profile 
convolved with the PSF. The dashed line is the $1\sigma$ sky-subtraction error for one thumbnail image, 
while the dash-dotted line is the $1\sigma$ sky-subtraction error for the composite stack of 48 LBGs,
as described in Section 4.2.
}
   \label{fig:sb}
\end{figure}

We extract a radial surface brightness profile from the super-stack of LBG images in Figure \ref{fig:sb} using custom IDL code which yields identical results to 
the IRAF\footnote{IRAF is distributed by the National Optical Astronomy Observatory, 
which is operated by the Association of Universities for Research in Astronomy,
Inc., under cooperative agreement with the National Science Foundation.} 
procedure $\tt ELLIPSE$. We use circular aperture rings with radial widths of 1.5 pixels in such a way 
that they do not overlap to avoid correlated data points while still finely probing the profile. 
We use custom code in order to facilitate using the bootstrap 
error analysis method to determine the uncertainties. 
Since we are stacking different objects with different characteristics, the uncertainty for the brighter regions will be dominated by the sample variance,
and can be determined with the bootstrap method. 
Specifically, we bootstrap to get the uncertainty of the median of the LBGs by replacing a random fraction ($1/e \approx 37\%$)
of the 48 LBG thumbnails with randomly duplicated thumbnails from the subset sample. We repeat this
1000 times to get a sample of composite images using the method described in Section 4.1, 
each with its own radial surface brightness profile. The resultant uncertainty is then the 
standard deviation of all the surface brightness magnitudes at each radius. This method
is conservative, but includes all the uncertainties associated with the variance of the sample. 
Moreover, it helps takes into account possible errors introduced by possible contamination of our LBG sample
by other objects, although as mentioned in Section 3, we believe this contamination fraction to be small. 

The solid blue line in Figure \ref{fig:sb} represents the measured ACS $V$ band
PSF determined from a stacked image of the 15 stars described in Section 3.2. 
The central surface brightness of the stars is scaled to match the LBGs. We note that the 
S/N of the star stack is higher than that of the LBG stack, even though it has a 
smaller number of objects in the stack, because the stars are brighter than the LBGs.
Therefore the PSF is well determined, and we expect the uncertainties are dominated
by the lower S/N LBG stack. 
The PSF declines more rapidly than the radial surface brightness profile at all radii, 
which shows that the LBGs are clearly resolved and that the median surface brightness profile 
of the LBGs is extended. The dashed line is the $1\sigma$ sky-subtraction error for one thumbnail image, 
while the dash-dotted line is the $1\sigma$ sky-subtraction error for the composite stack of 48 LBGs,
as described in Section 4.2. 

\subsubsection{S\'{e}rsic Profile Fit}

The dotted red line in Figure \ref{fig:sb} shows the best fit S\'{e}rsic profile convolved with the
ACS $V$ band PSF. We fit for the best S\'{e}rsic model to the composite LBG image 
by using the Levenberg--Marquardt least-squares minimization,
with the $\chi^2$ calculated on a pixel by pixel basis for each possible model.
The best fit values are $n=1.9\pm0.04$ and $R_e=0.074\pm0.001$ arcsec,
where $n$ is the S\'{e}rsic index and $R_e$ is the effective radius, 
which includes half the light of the LBGs. 
We use the dimensionless scale factor $b$($n$) from \citet{Ciotti:1999p20134}
such that $R_e$ is the half-light radius. 
The small $R_e$ value is likely due to our pre-selection of LBGs to be compact (see Section 3.2). 
While the fit has a good reduced chi-square 
$\chi^2/\nu$ of 1.27, it is not a very good fit to the data in the outer regions. 
The fit is dominated by the inner part of the 
profile with radii less than $\sim$0.{\arcsec}4, since the uncertainties are significantly smaller
in that region. For radii larger than $\sim$0.{\arcsec}4, the profile deviates from the inner best-fit profile.
This deviation is real, being above the PSF and
the $1\sigma$ sky-subtraction error. This is similar to what  
\citet{Hathi:2008p4334} found when stacking LBGs at $z\sim 4--6$. The main constraint we have from this
is that it suggests that the profile is similar to an exponential disk type profile. 
If we fit an exponential, it yields a worse fit with a $\chi^2/\nu$ of 1.4, and 
$R_e=0.068\pm0.001$ arcsec. A bulge-disk model yields a slightly better fit in the outer regions, 
but does not improve the overall $\chi^2/\nu$.
We note that we do not use the fit in the analysis below.

\subsection{Effects of the \lya Line on the Image Stack}

We investigate possible contamination from \lya emission on the image stack to ensure that the radial 
surface brightness profile is unaffected by \lya emission from the LBGs. First, we note that
\lya only enters our $V$ band filter for about half our sample due to the redshift range sampled. Second,
we find that due to the large width of the $V$ band filter, the \lya line would have a very small effect. This is
determined by taking the stacked LBG spectrum from \citet{Shapley:2003p4902}, and 
comparing the flux in the $V$ band filter with and without the \lya line. 
We find that the inclusion of the \lya line yields an increase of 
0.02 mag, a very small effect compared to our uncertainties.
While our sample of LBGs is significantly fainter than the LBGs in the stacked spectrum from 
\citet{Shapley:2003p4902}, we expect the average strength of the \lya line to be similar due
to the low escape fraction of ionizing radiation from LBGs \citep{Shapley:2006p3117}.

Moreover, we compare the radial surface brightness profile of the $V$ band stack
and an equivalent stack in the $i^\prime$ band.  The \lya line does not
enter the $i^\prime$ band filter throughout our redshift range, making the 
$i^\prime$ band an excellent test to check if the radial surface brightness profile
is affected by the \lya line.
We find that the $V$ and $i^\prime$ bands 
have the same radial surface brightness profile within their uncertainties,
with no systematic shifts.
This suggests that the \lya line 
has little to no effect, even if the \lya line were more extended than the continuum.

\section{Direct Inferences from the Data}

The surface brightness profile of the stacked image 
indicates the presence of spatially extended star formation around LBGs. 
In this section, we describe how to connect this emission,
which corresponds to the rest-frame FUV radiation intensity, to the SFR surface density. We then compare the covering fraction 
of this emission to that of the gas responsible for the star formation in Section 5.2. Specifically, in order to determine
what types of gas can be responsible for the observed emission, 
we compare its covering fraction to that of atomic-dominated gas in Section 5.2.1 and to that of molecular-dominated gas
in Section 5.2.2. Then in Section 5.3 and 5.4, we calculate the in situ SFR,  {\rhodot}, and metal production based on the 
integrated flux measured in the outskirts of LBGs. 

\subsection{Connecting the Observed Intensity to the SFR Surface Density}

In order to investigate how the measured star formation relates to the underlying gas, we require a relation
between star formation and gas properties. 
Star formation occurs in the presence of cold atomic and/or molecular gas,
according to the KS relation given by 

\begin{equation}
\Sigma_{\rm SFR} = K\times\left(\frac{\Sigma_{\rm gas}}{\Sigma_{\rm c}}\right)^\beta\;.
\label{eq:Kenrelation}
\end{equation}

\noindent This relation holds for nearby disk galaxies in which $\Sigma_{\rm gas}$ is
the mass surface density perpendicular to the plane of the disk, $\Sigma_{\rm c}=1$$M_\odot$pc$^{-2}$,
 $K=K_{\rm Kenn}$=(2.5$\pm$0.5){$\times$}10$^{-4}$ {\smpykpc}, 
 and ${\beta}$=1.4$\pm$0.15 \citep{Kennicutt:1998p3174, Kennicutt:1998p10742}.
There has been much recent work on improving both our understanding
of this relation, and measuring the values of $K$ and $\beta$
\citep[e.g.,][]{Leroy:2008p10197, Bigiel:2008p10261,Krumholz:2009p9796, Gnedin:2010p12943, 
Genzel:2010p19560, Bigiel:2010b}. 
When only considering molecular gas, the relationship has a flatter slope of 
 ${\beta}$=0.96$\pm$0.07 \citep{Bigiel:2008p10261} or ${\beta}\sim1.1$ \citep{Wong:2002p13491}.
We use the original values from \citep{Kennicutt:1998p3174} when considering the total gas density to 
simplify comparisons with other work, and ${\beta}$=1.0 and $K$=$K_{\rm Biegel}$=(8.7$\pm$1.5){$\times$}10$^{-4}$ {\smpykpc}
when considering only molecular gas. We note that the $K$ value given here used for molecular gas is modified from 
\citet{Bigiel:2008p10261} to use the same $\Sigma_{\rm c}=1~$$M_\odot$pc$^{-2}$ value as above. 

Rewriting the KS relation in terms of the column density of the gas, we get

\begin{equation}
\Sigma_{\rm SFR}=K\times\left(\frac{N}{N_c}\right)^{\beta},
\label{eq:KenrelationN}
\end{equation}

\noindent where the scale factor $N_{c}$=1.25{$\times$}10$^{20}$ cm$^{-2}$
\citep{Kennicutt:1998p3174,Kennicutt:1998p10742} and
$N$ is the hydrogen column density.\footnote{The reader should be aware that when 
referring to nearby galaxies, $N$ corresponds to $N_\perp$, the \ion{H}{1} column density 
perpendicular to the disk, but when writing about our observations, we are referring
to observed column densities $N$, where we implicitly include the inclination angles in
our definitions.} 
We note that this is only valid above the critical column density, which is usually 
associated with the threshold condition for Toomre instability. 
For \ion{H}{1} gas in local galaxies, it is observed to range between 5{$\times$}10$^{20}$ cm$^{-2}$
and 2{$\times$}10$^{21}$ cm$^{-2}$ \citep{Kennicutt:1998p10742}.

In order to connect $\Sigma_{\rm SFR}$ to the observations, we require a relation between observed intensity,
corresponding to rest-frame FUV emission, and observed column density, $N$. 
Following \citet[][equation (3)]{Wolfe:2006p474}, we find that for a fixed value of $N$, the intensity averaged
over all disk inclination angles is given by

\begin{equation}
\langle I_{\nu_0}^{\rm obs}\rangle=\frac{C\Sigma_{\rm SFR}}{4{\pi}(1+z)^{3}{\beta}}\;,
\label{eq:sbtoN}
\end{equation}

\noindent where $z$ is the redshift, and $C$ is the conversion factor from SFR to FUV
 ($\lambda\sim1500\AA$) radiation, with $C=8.4\times10^{-16}$ erg cm$^{-2}$
s$^{-1}$ Hz$^{-1}$({\smpykpc})$^{-1}$ \citep{Madau:1998p13049, Kennicutt:1998p10742}.
We use the same value of $C$ 
as \citet{Wolfe:2006p474} corresponding to a Salpeter IMF. 
The result in Equation (\ref{eq:sbtoN})
assumes that the star formation occurs in disks inclined on the 
plane of the sky by randomly selected inclination angles
and averages over all possible angles \citep[see][]{Wolfe:2006p474}.

\subsection{Covering Fraction of LBGs Compared to the Underlying Gas}

The covering fraction of observed star formation should be consistent with that of its underlying gas. 
We therefore investigate whether the covering fraction of the outer parts of LBGs is consistent with 
the covering fraction of atomic-dominated gas  in Section 5.2.1 and molecular-dominated gas in Section 5.2.2.
This consistency check yields insights into the nature of the observed star formation 
and validates the hypothesis that the outskirts of LBGs consist of atomic-dominated gas, 
which is used in subsequent sections of the paper.

We calculate the cumulative covering fraction, $C_A$, for gas columns greater than some 
column density, $N$, by integrating the H column-density distribution function $f(N_{\rm H},X)$,
where H is either \ion{H}{1}or H$_2$. Specifically, 

\begin{equation}
C_A(N) = {\int_{X_{\rm min}}^{X_{\rm max}}}dX {\int_{N}^{N_{\rm max}}dN_{\rm H} f(N_{\rm H},X)}\;,
\label{eq:covfrac}
\end{equation}

\noindent where $f(N_{\rm H},X)$ is the observed column-density distribution function of the hydrogen gas, 
$N_{\rm max}$ is the maximum column-density considered, 
and $X$ is the absorption distance with $dX$ being defined as

\begin{equation}
dX \equiv \frac{H_0}{H(z)}(1+z)^2dz\;,
\label{eq:X}
\end{equation}

\noindent where $H_0$ is the Hubble constant and $H(z)$ is the Hubble parameter at redshift $z$.
For $X_{\rm min}$ and $X_{\rm max}$ we use the same redshift interval as in Section 3.1, namely $2.7<z<3.4$
corresponding to $6.6<X<9.2$. 
The covering fraction depends strongly on the column-density distribution function, which is different for 
atomic and molecular gas. Below we investigate the covering fraction for both cases.

\subsubsection{Covering Fraction of Atomic-dominated Gas}

There is strong evidence to support the association of LBGs and neutral atomic-dominated \ion{H}{1} gas, i.e., DLAs (see Section 1), 
and we therefore investigate whether the covering fraction of the outer parts of LBGs is consistent with the covering 
fraction of DLAs. If the outer regions of LBGs truly consist of DLA gas, then the covering fractions as functions of the
surface brightnesses should be consistent. In this subsection, we work under the hypothesis that the observed FUV 
emission in the outskirts of LBGs is from in situ star formation in atomic-dominated gas and compare the 
observed covering fraction to that of the gas distribution.

In order to calculate the covering fraction using Equation (\ref{eq:covfrac}),
we require $f(N_{\rm H},X)$ for atomic-dominated gas. 
The observed \ion{H}{1} column-density distribution function, $f(N_{{\rm H  I}},X)$, 
is obtained by a double power-law fit to the Sloan Digital Sky Survey data:

\begin{equation}
f(N_{{\rm H  I}},X)=k_{3}\left(\frac{N}{N_{0}}\right)^{\alpha}\;,
\label{eq:fofN}
\end{equation}

\noindent where $k_{3}$=(1.12$\pm$0.05)$\times$10$^{-24}$ cm$^{2}$,
$\alpha$=$\alpha_{3}$=$-$2.00$\pm$0.05 for $N\le N_0$\footnote{
We follow \citet{Wolfe:2006p474} who equated $N_0$ with $N_d$, 
the break in the double power-law expression for $f(N_{{\rm H  I}},X)$.}
 and $\alpha$=$\alpha_{4}$=$-$3.0 for $N>N_0$, where
 $N_0=3.54^{+0.34}_{-0.24} \times 10^{21}$ cm$^{-2}$
 \citep{Prochaska:2005p479, Prochaska:2009p11257}. 
 The value of $\alpha_{4}$ used is different than measured in
\citep{Prochaska:2009p11257}, to remain consistent with 
our formulation of randomly oriented disks in Section 6, 
and is predicted to be $-3.0$, and we use this value for the covering fraction to be consistent.
Although this value of $\alpha_{4}$ is different than the value in \citet{Prochaska:2009p11257},
the uncertainties are quite large due to low numbers of very high column density DLAs, 
and it is quite similar to the value found by \citet{Noterdaeme:2009p11271} of $\alpha_{4}$=$-$3.48.

\begin{figure}[t!]
\center{
\includegraphics[scale=0.5, viewport=15 5 495 350,clip]{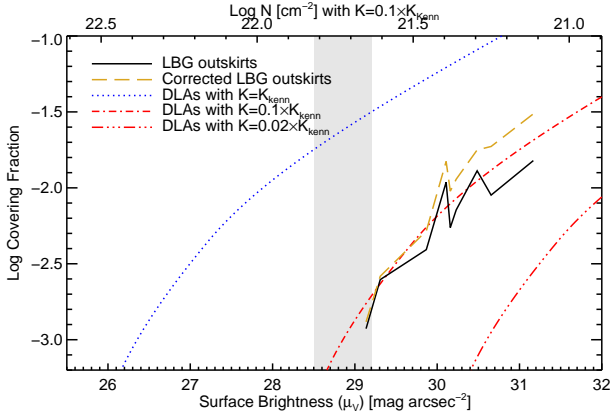}}
\caption{Cumulative covering fraction of DLA systems with columns greater than some column density $N$ (and therefore surface brightness), 
compared to the covering fraction of the observed $z\sim3$ LBG outskirts greater than some surface brightness. 
The blue dotted line is the DLA covering fraction for DLAs with $K=K_{\rm Kenn}$, the red dash-dotted line is for $K=0.1\times K_{\rm Kenn}$,
and the red triple-dot-dashed line is for $K=0.02\times K_{\rm Kenn}$.  
The solid black line is the covering fraction for the LBG outskirts, while the gold long-dashed line is the same corrected for completeness.
The gray filled region represents the transition region mentioned in Section 5.2.2. 
The top $x$-axis coordinates are the column densities corresponding to the surface brightnesses using an efficiency of $K=0.1\times K_{\rm Kenn}$.
}
\label{fig:cov_frac}
\end{figure}

We note that  \citet{Noterdaeme:2009p11271} find slightly different values for  $k_{3}$ and $\alpha_{3}$
than \citet{Prochaska:2009p11257}, with
 $k_{3}$=8.1$\times$10$^{-24}$ cm$^{2}$ and
$\alpha$=$\alpha_{3}$=$-$1.60 for $N\le N_0$, 
corresponding to a flatter slope. We use the values from \citet{Prochaska:2009p11257}
and describe how the differing values affect our results below. 
Also, although the normalization of $f(N_{{\rm H  I}},X)$ varies with redshift, the variations for 
our redshift interval are not large and do not strongly affect $C_A$. 
 Using the \citet{Prochaska:2009p11257} values for
$f(N_{{\rm H  I}},X)$, and $N_{\rm max} = 10^{22}$ cm$^{-2}$, we calculate the covering fraction
using Equation (\ref{eq:covfrac})
and the corresponding expected $\mu_V$ from Equations (\ref{eq:KenrelationN}) and (\ref{eq:sbtoN}),
where $\mu_V=-2.5{\rm log}\langle I_{\nu_0}^{\rm obs}\rangle-Z$, and $Z$ is the AB magnitude zero point of 26.486.

We compare the cumulative covering fraction of DLAs to the observed covering fraction of the outer regions
of the LBGs in Figure \ref{fig:cov_frac}. The blue dotted line is the DLA covering fraction for
DLAs forming stars according to the KS relation, with $K=K_{\rm Kenn}$. The red
lines are for less efficient star formation where the dash-dotted line represents $K=0.1\times K_{\rm Kenn}$
and the triple-dott-dashed line represents $K=0.02\times K_{\rm Kenn}$. 
The black line is the covering fraction for the observed emission in the outskirts of LBGs in the UDF,
which is just the area covered by the outskirts of LBGs up to $\mu_V$ 
divided by the total area probed, 11.56 arcmin$^{2}$.
The covered area is obtained from the radial surface brightness profile and the 407 observed
LBGs in this area. 
The observed covering fraction depends on the depth of the images and therefore 
requires a completeness correction for faint objects that are missed: we discuss this
completeness correction in Appendix A. The completeness 
corrected covering fraction for LBGs is the dashed gold long-dashed line, and is 
the curve that we compare to the DLA lines below. 
We ignore the data at a radius 
$>0.8$ arcsec ($\sim6$kpc) to only include data with S/N$>$3, however, we expect a continuation
of the observed trends. 
We note that we are only sampling the top end of the DLA
distribution function, and therefore the covering fraction shown is a small fraction (about a tenth) of
the total covering fraction of DLAs. When considering all DLAs, we cover about one-third of the sky,
i.e., ${\rm log}(C_A) \sim -0.5$.
We note that if we use the \citet{Noterdaeme:2009p11271} values for $f(N_{{\rm H  I}},X)$, 
then the DLA lines move up and to the left (i.e., cover more of the sky).

Under the hypothesis that the outskirts of LBGs are comprised of atomic-dominated gas (i.e., DLAs) which is responsible for the 
observed emission, then 
we expect the covering fraction of the DLA gas to be 
equal to the covering fraction of the outskirts of LBGs.
The blue dotted line for DLAs following the KS relation would therefore
only be consistent with the covering fraction of the outskirts of LBGs if much of the DLA
gas is not surrounding LBGs. This possibility was constrained by \citep{Wolfe:2006p474},
who found that DLAs would need to be forming at significantly lower SFR efficiencies if this
was the case. 
On the other hand, we find that the covering fraction of the outskirts of LBGs is consistent with
DLAs having SFR efficiencies of $K\gtrsim0.1\times K_{\rm Kenn}$, at which point the covering fraction
is roughly equal. This covering-factor analysis provides evidence that
if the outskirts of LBGs are comprised of DLA gas, then the SFR efficiency
of this atomic-dominated gas at $z\sim3$ is about 10\% of the efficiency for local galaxies.

Moreover, the cumulative covering fraction shows that there is sufficient DLA gas available to be
responsible for the emission in the outskirts of LBGs for SFR efficiencies $\gtrsim10$\%.
We investigate this lower SFR efficiency further in Section 6, where we find an efficiency 
closer to 5\%, which is only a factor of $\sim$2 different than the efficiency determined from the covering
fraction. We note that systematic uncertainties due to assumptions throughout both quantities 
could easily be off by a factor of two, and so the general
agreement of the covering fraction at low SFR efficiencies is reassuring, 
and we are not concerned about the minor disagreement.

\subsubsection{Covering Fraction of Molecular Gas}

In the previous subsection, we worked under the hypothesis that the observed FUV emission in the outskirts of LBGs is from
in situ star formation in atomic-dominated gas. However, it is possible that this star formation occurs in 
molecular-dominated gas. We consider this scenario here, and compare the covering fraction of molecular-dominated gas,
where the majority of the hydrogen gas is molecular, to our observations.

In order to calculate the covering fraction using Equation (\ref{eq:covfrac}), we require $f(N_{\rm H},X)$ for molecular-dominated gas. 
We use the observed molecular column-density distribution function, $f(N_{\rm H_2})$,  from \citet{Zwaan:2006p13760},
who obtained  a lognormal fit to the BIMA SONG data \citep{Helfer:2003p18674}, 

\begin{equation}
f(N_{\rm H_2})=f^*~{\rm exp}\left[\left(\frac{{\rm log}N-\mu}{\sigma}\right)^2 /~2\right]\;,
\label{eq:lognormal}
\end{equation}

\noindent where $\mu=20.6$, $\sigma=0.65$, and the normalization 
$f^*$ is $1.1\times10^{-25}$ cm$^2$ \citep{Zwaan:2006p13760}\footnote{We note
that \citet{Zwaan:2006p13760} have a typographical error, switching $\mu$ and $\sigma$.}.
We use the molecular version of the KS relation discussed in Section 5, where
${\beta}$=1.0 and $K$=$K_{\rm Biegel}$=8.7{$\times$}10$^{-4}$ {\smpykpc}, 
and we let $N_{\rm max} = 10^{24}$ cm$^{-2}$, the largest observed value for the $f(N_{\rm H_2})$ 
function used \citep{Zwaan:2006p13760}. 
However, $f(N_{\rm H_2})$ is not observationally determined at $z\sim3$, and likely evolves over time, and we investigate such a possibility below. 
 
The evolution of $f(N_{\rm H_2})$ would either be due to a change in the normalization or a change in the shape.
The shape of the atomic gas column-density distribution function, $f(N_{\rm H_I})$, has not evolved 
between $z=0$ and $z=3$ \citep{Zwaan:2005p15438, Prochaska:2005p479, Prochaska:2009p11257},
but the normalization has increased by a factor of two. 
In the case of H$_2$, we consider the instance in which only the normalization ($f^*$)
evolves, then we are looking for a change in $\Omega_{\rm H_2}$ at $z=0$ to $\Omega_{\rm H_2}$ at $z=3$.
While theoretical models predict that $\Omega_{\rm H_2}(z=3)/\Omega_{\rm H_2}(z=0)$ is $\sim4$,
their similar prediction for atomic gas does not match observations \citep{Obreschkow:2009p14984}.
Alternatively, we can determine an upper limit of the evolution of $f^*$ using the evolution of {\rhodot} for galaxies between $z=0$ and $z=3$, 
assuming that the evolution in $f(N_{\rm H_2})$ is only due to the normalization, and there is no evolution in the KS law for molecular gas between $z=0$ and $z=3$ 
\citep[][see Section 6.3]{Bouche:2007p4658, Daddi:2010p13822,Genzel:2010p19560}. Specifically,  studies have found that
 {\rhodot}$(z=3)$/{\rhodot}$(z=0)\sim10$ \citep{Schiminovich:2005p15693, Reddy:2008p4837}, and therefore $f^*$
 changes at most by a factor of 10, if we assume that the contribution from atomic gas is small. This is used as 
 an upper limit to the evolution of $f^*$, and we are not suggesting that this is the correct evolution. 
  
\begin{figure}[t]
\center{
\includegraphics[scale=0.5, viewport=15 5 495 350,clip]{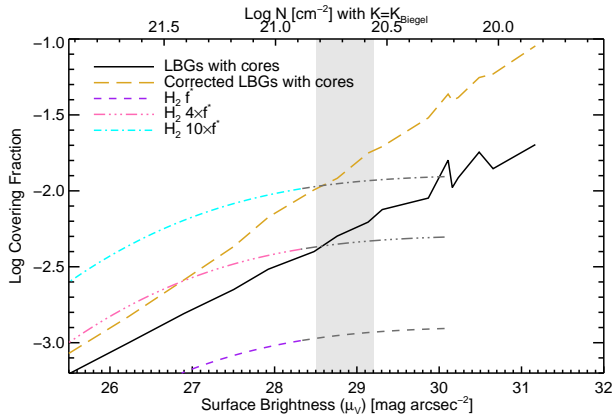}}
\caption{Cumulative covering fraction of gas with columns greater than some column density $N$ (and therefore surface brightness). 
This figure is similar to Figure \ref{fig:cov_frac}, but makes comparisons to the covering fractions of molecular gas rather than atomic-dominated gas. 
The solid black line is the covering fraction for LBGs starting from the center of the LBG core, and the gold long-dashed line
is the same corrected for completeness. The purple short-dashed line is the covering fraction of molecular hydrogen with no evolution at $z\sim3$,
the pink triple dot-dashed line is the same with an evolution of four times $f^*$ 
(the normalization of the column-density distribution function of molecular gas)
and the cyan dot-dashed line is the same with an evolution of 10 times $f^*$.
The gray lines continuing the purple, pink, and cyan lines are extrapolations of the data to lower column densities. 
The gray filled region represents the transition region mentioned in Section 5.2.2. 
The top $x$-axis coordinates are the column densities corresponding to the surface brightnesses 
using an efficiency of $K=K_{\rm Biegel}$ and $\beta =1$, valid for molecular gas. 
}
\label{fig:cov_frac_h2}
\end{figure}

We plot the covering fractions of molecular gas at $z=3$ 
in Figure \ref{fig:cov_frac_h2}.
We consider three cases for $f(N_{\rm H_2})$:
(1) no evolution as a purple short-dashed line, 
(2) evolution with a factor of four increase in $f^*$
based on the model by \citet{Obreschkow:2009p14984} as a pink triple-dashed line, and
(3) evolution with a factor of 10 increase in $f^*$ based on the observed evolution in {\rhodot}  as a cyan dot-dashed line. 
The gray lines continuing these three lines are extrapolations of the data to lower column densities than observed.
We note that the column densities on the top $x$-axis of the plot 
now are for $K=K_{\rm Biegel}$.

These covering fractions of molecular gas are compared to the LBG profile including the inner cores in Figure \ref{fig:cov_frac_h2}. 
{The LBG covering fraction is now modified to include the LBG cores which are composed of molecular-dominated gas, and 
the new LBG covering fraction is plotted as a solid brown line. 
We again ignore the data at a radius 
$>0.8$ arcsec ($\sim6$kpc) to only include data with S/N$>$3.
This covering fraction is also corrected for completeness similar to Section 5.2.1, except this time we include the cores of the LBGs 
and make no distinction between the outskirts and the inner parts of the LBGs, and is described in Appendix A.1.
We plot this corrected covering fraction as a 
gold long-dashed line. This correction is quite large, as even though the cores cover a smaller area than the 
outskirts, the cores of fainter missed LBGs contribute at every surface brightness as we go fainter. 
Since there are significantly more faint LBGs than bright LBGs,  there are significantly more LBG cores contributing to each surface 
brightness than there are LBGs with outskirts at those same surface brightnesses. This correction assumes
that all the star formation comes from molecular-dominated gas, and therefore all the cores of fainter LBGs are included.

As in Section 5.2.1, we expect the covering fraction of the gas to be equal to the 
covering fraction of the LBGs forming out of that gas. In the case of purely molecular gas,
the upper limit of the covering fraction ($10\times f^*$) of the gas is only consistent to
$\mu_V\sim28.5$. At $\mu_V \gtrsim28.5$, the covering fraction of LBGs is larger than that of the 
molecular gas. In order to have a covering fraction larger than all the corrected LBG emission,
we would require an evolution of $f^*$ by a factor of more than 60, which is not very likely.

We have just shown that the predicted covering factor for
molecules is inconsistent with our results for $\mu_V \gtrsim28.5$.
At the same time, our results for atomic-dominated hydrogen
do not apply for $\mu_V$ brighter than $\sim29$, due to the 
atomic-dominated \ion{H}{1} gas cutoff of $N_{\rm HI} \le 10^{22}$cm$^{-2}$. Therefore, 
the surface brightness interval from 28.5 $\lesssim \mu_V \lesssim$ 29.2 is not simultaneously 
consistent with the neutral atomic-dominated gas nor the molecular gas' covering fractions. 
This lower bound on $\mu_V$ is based on comparing the LBG data compared to 10 times $f^*$, 
and for lower evolutions of  $f^*$, this begins at even brighter $\mu_V$.
This result is reasonable if the LBG 
outskirts consist of atomic-dominated gas. In this scenario, there would need to be 
a transition region between atomic-dominated and molecular-dominated gas, where
star formation occurs in both phases. We note that in this hybrid region, the corrected covering 
fraction of LBGs is not correct, as we would be adding the star formation in the cores of 
missed LBGs to the outskirts presumably composed of atomic-dominated gas.
The true correction would lie somewhere
between the black solid line and the gold long-dashed line.
We take the molecular gas covering 
fraction results as evidence that the outskirts of LBGs consist of atomic-dominated
gas, which is consistent with our underlying hypothesis for this paper.

\begin{deluxetable*}{llcrrrrrr}
\tabletypesize{\scriptsize}
\tablecaption{SFR, $\dot{\rho_{*}}$, and Metallicity
\label{tab2}}
\tablewidth{0pt}
\tablehead{
\colhead{$\theta_{\rm low}$\tablenotemark{a}} &
\colhead{$\theta_{\rm high}$\tablenotemark{a}} &  
\colhead{$R/H$\tablenotemark{b}} & 
\colhead{$\rm SFR$} &
\colhead{$\dot{\rho_{*}}$}   & 
\colhead{$\dot{\rho_{*}}$ Corrected} &
\colhead{$f$\tablenotemark{c}} &
\colhead{$Z$} &
\colhead{$\rm [M/H]$} \\
\colhead{(arcsec)} &
\colhead{(arcsec)} &
\colhead{} &
\colhead{($M_\odot$ yr$^{-1}$) }  &
\colhead{($M_\odot$ yr$^{-1}$ Mpc$^{-3}$) } &
\colhead{($M_\odot$ yr$^{-1}$ Mpc$^{-3}$) } &
\colhead{} &
\colhead{$(Z_{\odot}$)} &
\colhead{}
}

\startdata

0.405 &0.765 &10 &$0.118\pm 0.002$  &$(1.82\pm0.03)\times10^{-3}$ &$(3.75\pm0.03)\times10^{-3}$ &$0.084\pm0.001$ &$0.12\pm0.05$ &$-0.9\pm0.4$  \\
0.405 &0.765 &100 &$0.059\pm0.001$ &$(0.91\pm0.01)\times10^{-3}$ &$(1.87\pm0.01)\times10^{-3}$ &$0.084\pm0.001$ &$0.12\pm0.05$ &$-0.9\pm0.4$  \\
0.405 &1.125 &10 & $0.187\pm 0.005$ &$(2.87\pm0.08)\times10^{-3}$ &$(5.91\pm0.08)\times10^{-3}$ &$0.126\pm0.002$ &$0.19\pm0.07$ &$-0.7\pm0.4$  \\
0.405 &1.125 &100 &$0.093\pm0.002$ &$(1.44\pm0.04)\times10^{-3}$ &$(2.95\pm0.04)\times10^{-3}$ &$0.126\pm0.002$ &$0.19\pm0.07$ &$-0.7\pm0.4$
\enddata

\tablecomments{The integrated SFR and $\dot{\rho_{*}}$ in the outskirts of LBGs. We integrate from 
the point where the theoretical models for DLA gas and the LBG data overlap, in order
to probe the hypothesis that the FUV emission in this region is from in situ star formation
in DLA gas associated with the LBGs.}

\tablenotetext{a}{Radii from the center of the composite LBG stack.}
\tablenotetext{b}{The thickness of the disk, where $R$ is the radius and $H$ is the scale height.}
\tablenotetext{c}{Fraction of {\rhodot} observed in the outer region of the composite LBG stack divided by the total {\rhodot} observed.}

\end{deluxetable*}

\subsection{Measurements of the SFR and {\rhodot} in the Outskirts of LBGs}

In addition to calculating the efficiency of the SFR, we can also 
calculate the mean SFR,  $\langle {\rm SFR}\rangle$, and {\rhodot} in the outskirts of LBGs by integrating the
rest-frame FUV emission in the outer areas. These measurements allow us to 
calculate the metals produced in the outskirts of LBGs in Section 5.4 and to put a limit
on the total {\rhodot} contributed by DLA gas. 
Similar to \citet{Wolfe:2006p474}, we assume that DLAs are disk like structures, 
or any other type of gaseous configurations with preferred
planes of symmetry. We note that while we work this out for DLA gas, the only assumption in
our derivation is that the outskirts have preferred planes of symmetry such, as disks. 
Even if the outskirts of LBGs are not DLA gas, such an assumption is still valid given the rotation curves
measured for high redshift LBGs \citep[e.g.,][]{ForsterSchreiber:2009p11349}
and the predictions by simulations \citep{Brooks:2009p16590, Ceverino:2010p21099}.

For these calculations, 
we need  $\Sigma_\nu^\perp$, the luminosity per unit frequency interval per unit area 
projected perpendicular to the plane of the disk. 
Specifically, we solve for  $\Sigma_\nu^\perp$ 
as a function of $\langle I_{\nu_0}^{\rm obs}\rangle$ 
averaged over all inclination angles. We find that 

 \begin{equation}
\Sigma_\nu^\perp \equiv \frac{4\pi(1+z)^3\langle I_{\nu_0}^{\rm obs}\rangle}{\ln(R/H)}\;,
\label{eq:Sigmanu}
\end{equation}

\noindent where $R$ is the radius and $H$ is the scale height of the model disks, 
which holds in the limit $R\gg H$.  We calculate $\Sigma_\nu^\perp$
for a range in aspect ratios, with $R/H$ values from 10 to 100, covering a range from thick disks, 
as possibly seen at high redshift \citep[e.g.,][]{ForsterSchreiber:2009p11349},  all
the way to thin disks resembling the Milky Way. We then calculate the mean SFR by integrating  
$\Sigma_\nu^\perp$ across the outer region of 
the LBG stack, and find that

\begin{equation}
\langle {\rm SFR} \rangle = \frac{8\pi(1+z)^3}{C' {\rm ln}(R/H)} \int_{\theta_{\rm low}}^{\theta_{\rm high}}2\pi {d_A}^2 \langle I_{\nu_0}^{\rm obs}(\theta)\rangle \theta d\theta \;,
\label{eq:SigmaSFR}
\end{equation}

\noindent where $d_A$ is the angular diameter distance, $\theta$ is the radius in arcseconds, $\theta_{\rm low}$ is the minimum radius
for the outer region, and $\theta_{\rm high}$ is the maximum radius. 
$I_{\nu_0}^{\rm obs}(\theta)$ comes from the radial surface brightness profile 
from Section 4.3 and depends on $\mu_V(\theta)$, namely
$I_{\nu_0}^{\rm obs}(\theta) = 10^{-0.4(\mu_V(\theta)+48.6)}$. 
The SFR depends on the inclination angles of the disks for a given $I_{\nu_0}^{\rm obs}$
and includes a factor of two for averaging over all inclination angles.

In order to find the $\langle {\rm SFR} \rangle$ in the outer regions of LBGs, we need to designate 
a radius at which to start integrating 
the LBG stack,
and a comparison of the theoretical model to the data in Section 6.2 yields this radius. 
The $\langle {\rm SFR} \rangle$ is independent of the theoretical framework developed later in Section 6
and does not depend on the efficiency of the gas. It is purely a measurement 
of the star formation occurring in the gas. However, it requires a minimum radius to 
define the beginning of the outer region of the LBGs. 
Specifically, we pick the smallest radius that corresponds to the first point in Figure \ref{fig:rhodotstar_sb_comp}
where we demonstrate that the smallest radius of atomic-dominated gas corresponds to  0.{\arcsec}4.\footnote{
This corresponds to $\mu_V=28.8\pm0.2$ mag arcsec$^{-2}$, or 
$\Sigma_{\rm SFR} = (6 \pm 1) \times 10^{-3}$ $M_\odot$ yr$^{-1}$ kpc$^{-2}$ (Figure \ref{fig:sb}).}
We note that when we refer to spatially extended emission star formation 
throughout the paper, we are referring to emission at radii larger than 0.{\arcsec}4.
We also require a second
point that we integrate out to, for which we use two different values. First, we use
the radius of 0.{\arcsec}8\footnote{This corresponds to
$\mu_V=30.5\pm0.5$ mag arcsec$^{-2}$, or 
$\Sigma_{\rm SFR} =(1.3\pm0.6) \times 10^{-3} $ $M_\odot$ yr$^{-1}$ kpc$^{-2}$(Figure \ref{fig:sb}).}, 
which corresponds to the largest radius above
3$\sigma$. Second, we integrate to 1.{\arcsec}1 
\footnote{This corresponds to
$\mu_V=30.6\pm0.6$ mag arcsec$^{-2}$, or 
$\Sigma_{\rm SFR} = (1.2\pm0.7) \times 10^{-3} $ $M_\odot$ yr$^{-1}$ kpc$^{-2}$(Figure \ref{fig:sb}).}, 
corresponding to the furthest point
for which we measured $\mu_V$ in Figure \ref{fig:sb}. 
Table \ref{tab2} lists the SFRs for different combinations of $R/H$ 
and $\theta_{\rm high}$. Since we integrate over the radii
where the LBG data intersect the theoretical models for the DLA gas, 
the FUV emission from this region may be 
from the in situ star formation in DLA gas associated with the LBGs (see Section 7). 

After we have the SFR we can calculate values for
the SFR per unit comoving volume, 
{\rhodot}, via {\rhodot}=SFR$\times N_{\rm LBG}/V_{\rm UDF}$,
and they are tabulated in Table \ref{tab2}. 
However, since {\rhodot} depends on $N_{\rm LBG}$,  we perform a completeness correction as described
in Appendix A.2. 
The resultant completeness corrected {\rhodot}  are listed in Table  \ref{tab2}. 
While there is a range in the acceptable values for both the SFR and {\rhodot}, we find that the 
extended emission has $\langle {\rm SFR} \rangle$ $\sim 0.1$ $M_\odot$ yr$^{-1}$ and 
{\rhodot} $\sim3\times10^{-3}$ $M_\odot$ yr$^{-1}$ Mpc$^{-3}$.

We take this measured {\rhodot}  in conjunction with the upper limit found in
\citet{Wolfe:2006p474} to calculate the total {\rhodot} from neutral atomic-dominated
gas at $z\sim3$. Specifically,  \citet{Wolfe:2006p474}  constrain {\rhodot} 
for regions in the UDF without LBGs, which complements the results
from this study for regions containing such objects. Together, we constrain 
all possibilities for the star formation from such gas. 
\citet{Wolfe:2006p474} place a conservative upper limit on {\rhodot}
contributed by DLAs with column densities greater than $N_{\rm min}=2\times10^{20}$cm$^{-2}$, 
finding {\rhodot} $<$ $4.0\times10^{-3}$ $M_\odot$ yr$^{-1}$ Mpc$^{-3}$. Combining this with 
 our largest possible value of the completeness corrected {\rhodot} in Table \ref{tab2} of 
{\rhodot}=$5.91\times10^{-3}$ $M_\odot$ yr$^{-1}$ Mpc$^{-3}$, we calculate an upper limit
on the total {\rhodot} contributed by DLA gas. 
We find a conservative upper limit  of {\rhodot}$<$ $9.9\times10^{-3}$ $M_\odot$ yr$^{-1}$ Mpc$^{-3}$,
corresponding to $\sim10\%$ of the {\rhodot} measured in the inner regions of 
LBGs at $z\sim3$ \citep{Reddy:2008p4837}.

\subsection{Metal Production in the Outskirts of LBGs}
 
Under the hypothesis that the outskirts of LBGs are composed of atomic-dominated gas,
we can calculate the metals produced due to in situ star formation from $z=10$ to $z=3$, 
and compare this to the metals observed in DLAs at $z=3$. 
The metal production can be measured from the outskirts of the LBG composite since the
FUV luminosity is a sensitive measure of star formation, since the massive stars
produce the UV photons as well as the majority of the metals. 
The comoving density of metals produced is obtained by integrating the 
comoving SFR density ({\rhodot}) from the most recent galaxy surveys 
\citep{Bouwens:2010p20181, Bouwens:2010p27816, Reddy:2009p6997}. 
We note that the resultant metallicities are only valid if the outskirts of LBGs 
are composed of atomic-dominated gas, as we divide by the 
H~I  mass density, $\rho_{\rm HI}$, to obtain the metallicity. 

First, we integrate {\rhodot} for all LBGs from $z\sim3$ to $z\sim10$ using the 
{\rhodot} values from \citet{Bouwens:2010p20181, Bouwens:2010p27816} and \citet{Reddy:2009p6997} to calculate
the total mass of metals produced in LBGs by $z\sim3$ similar to \citet{Pettini:1999p14867, pett04, Pettini:2006p11353} and \citet{Wolfe:2003p2460}.
Specifically, we calculate the comoving mass density of stars at $z\sim3$ by 

\begin{equation}
\rho_{*, {\rm LBG}} = \int^{z=10}_{z=3} \dot{\rho}_{*, {\rm LBG}} \frac{dt}{dz} dz = 1.1 \times10^8~M_{\odot}~{\rm Mpc^{-3}}
\label{eq:rhostar}
\end{equation}

\noindent where 

\begin{equation}
\frac{dt}{dz} = \frac{1}{(1+z) H(z)} \;.
\label{eq:dtdz}
\end{equation}

In order to obtain the comoving mass density of stars in the outskirts of LBGs, 
we multiply this result by the fraction of {\rhodot} observed in the outer region of the composite LBG stack
compared to the total {\rhodot} observed, $f$, which we list in Table \ref{tab2}. 
We can then calculate the total mass in metals produced by $z\sim3$ using the estimated conversion factor 
$\dot{\rho}_{\rm metals} = (1/64)~\dot{\rho_*}$ by \citet{Conti:2003p15049}, which is a factor of 1.5 lower than
the metal production rate originally estimated by \citet{Madau:1996p4663}. 
The metallicity of the presumed DLA gas is then calculated by dividing by $\rho_{\rm HI}$ at $z\sim3$, 
where we use average value of $\rho_{\rm HI}$ over the redshift range $2.4 \lesssim z \lesssim 3.5$ of
 $(9.0 \pm 0.1) \times 10^{7}$ $M_\odot$ Mpc$^{-3}$ \citep{Prochaska:2009p11257}.
 The final metallicities are tabulated in Table \ref{tab2} in terms of the solar metallicity,
where $Z_\odot$ = 0.0134 \citep{Asplund:2009p14796,Grevesse:2010p14713}. The metallicities
range from $ 0.12 Z_\odot$ to $ 0.19 Z_\odot$, similar to DLA metallicities (see Section 7.6). 
We note that $f$ is independent of the disk aspect ratio ($R/H$), 
and therefore so is the metallicity.

\section{Star Formation Rate Efficiency in Neutral Atomic-dominated Gas}

In our search for spatially extended LSB emission around LBGs, we aim to 
further our understanding of the connection between the DLA gas studied in absorption and the
star formation needed to explain the characteristics of the DLA gas. Specifically, 
 \citet{Wolfe:2006p474} searched for isolated LSB emission of DLAs, away from known LBGs, in the UDF
and found conservative upper limits on the SFR per unit comoving volume, {\rhodot}.  These limits constrain the in situ SFR 
efficiency of DLAs to be less than 5$\%$ of that expected from the KS relation. In other words, star formation must occur 
at much lower efficiency in neutral atomic-dominated hydrogen gas at $z\sim3$ than in modern galaxies at $z=0$. 

The surface brightness profile of the super-stack of 48 resolved LBGs 
(Figure \ref{fig:sb}) reveals the presence of spatially extended star formation around LBGs. 
The latest evidence suggests that this star formation is most likely occurring in atomic-dominated gas.
The most convincing evidence is measurements probing the outer disks of local galaxies that detect star formation
in atomic-dominated hydrogen gas \citep{Fumagalli:2008p5770, Bigiel:2010b, Bigiel:2010a}. 
In addition, 
we find that because the covering fraction of molecular gas is insufficient to explain the observed star formation in the outskirts of LBGs, 
the observed emission is likely from atomic-dominated gas (see Section 5.2).
Throughout the rest of this investigation, we work under the hypothesis that the observed 
FUV emission in the outskirts of LBGs is from 
in situ star formation in atomic-dominated gas.
In order to  quantify the efficiency of star formation at high redshift in atomic-dominated gas,
we require a theoretical framework connecting the observed emission around LBGs to the expectations based on known DLA statistics..

We develop such a framework in Section 6.1, where we combine the column-density distribution function of DLAs
with the KS relation to construct a model that predicts the comoving SFR density per intensity for different
SFR efficiencies of the KS relation. We convert the measured radial surface brightness profile from Section 4.3
into this same quantity in Section 6.2 and compare it to the model.  Through this comparison, we obtain an SFR efficiency for each surface 
brightness in the profile which corresponds to both a specific radius in the profile and a gas column density via 
the KS relation. 
As a tool to understand the SFR efficiencies and compare our results to those of \citet{Wolfe:2006p474} and \citet{Bigiel:2010b}, and simulations such as \citet{Gnedin:2010p12943},
we convert our results to fit onto a standard plot of $\Sigma_{\rm SFR}$ versus $\Sigma_{\rm gas}$ in Section 6.3. The
$\Sigma_{\rm SFR}$ is determined directly from the measured rest-frame FUV flux, and the $\Sigma_{\rm gas}$
is determined from the $\Sigma_{\rm gas}$, the KS relation, and the SFR efficiency determined through comparisons
of the data with the column-density distribution function. 

\subsection{Theoretical Framework}

We require a theoretical framework for the rest-frame FUV emission from DLAs and
start with the one developed in \citet{Wolfe:2006p474} for the expected emission from DLAs
in the $V$ band image of the UDF. After summarizing this framework, we expand it to explain
the observed emission around LBGs as a function of radius (and therefore surface brightness),
taking into account the projection effects of randomly inclined disks. Our resultant model in 
\S6.1.2 yields predictions of the differential {\rhodotstar} per intensity interval expected from DLAs for different 
SFR efficiencies. We then compare this model to the data in Section 6.2 to obtain the SFR efficiency
of the DLA gas. 

\subsubsection{Original Framework from \citet{Wolfe:2006p474} }

The framework developed in \citet{Wolfe:2006p474} connects the 
measured column density 
distribution function, $f(N_{{\rm H  I}},X)$, the KS relation, and randomly inclined disks
 to determine the expected cumulative
{\rhodotstar} for DLAs as a function of column density and therefore surface brightness. 
Specifically, they develop an expression for {\rhodotstar}
due to DLAs with observed column density greater or equal to $N$, and we take this expression directly
from Equation (6), in \citet{Wolfe:2006p474}, namely

\begin{equation}
\dot{\rho_*}({\ge}N,X)=\left(\frac{H_{0}}{c}\right) {\int_{N}^{N_{\rm max}}dN'J(N')\Sigma_{\rm SFR}(N')}\;.
\label{eq:rhostardotN}
\end{equation}

\noindent Here $H_0$ is the Hubble constant, $c$ is the speed of light, $N_{\rm max}$ is the maximum observed column
density for DLAs (10$^{22}$ cm$^{-2}$), and $J(N')$ is

\begin{equation}
J(N')={{\int_{N_{\rm min}}^{{\rm min}(N_{0},N')}dN_{\perp}g(N_{\perp},X)}}
\left(\frac{N_{\perp}^{2}}{N'^{3}}\right)\left(\frac{N_{\perp}}{N'}\right)^{{\beta}-1} \;\;\;.
\label{eq:JofN}
\end{equation}

\noindent Here $X(z)$ is the absorption distance and 
$g(N_{\perp},X)$ is the intrinsic column-density distribution of the disk
for which the maximum value of $N_\perp$, the \ion{H}{1} column density 
perpendicular to the disk, is $N_0$. $g(N_{\perp},X)$ is
related to the observed \ion{H}{1} column-density distribution function 
$f_{\rm H I}(N,X)$ by

\begin{equation}
f_{H\rm I}(N,X)={{\int_{N_{\rm min}}^{{\rm min}(N_{0},N)}dN_{\perp}g(N_{\perp},X)}}(N_{\perp}^{2}/N^{3}) 
\label{eq:fNNperp}
\end{equation}

\noindent \citep{Fall:1993p14497, Wolfe:1995p14525}.

A potential problem with using $f(N_{{\rm H  I}},X)$ in the expression
for {\rhodotstar} is that the measurements of $f(N_{{\rm H  I}},X)$ originate from
absorption-line measurements that sample scales of 
$\sim1$ pc \citep{Lanzetta:2002p17035}. On the other hand, 
the KS relation is established on scales exceeding 0.3 kpc \citep{Kennicutt:2007p4941}.
This is not an issue, however, because $f(N_{{\rm H  I}},X)$ typically depends on
over 50 measurements per column-density bin and is therefore a
statistical average over probed areas that exceed a few kpc$^2$ \citep[see][]{Wolfe:2006p474}.

\subsubsection{New Differential Approach for LBG Outskirts}

The framework developed in \citet{Wolfe:2006p474} was appropriate for connecting
the upper limit measurements of {\rhodot} from DLAs above a limiting column density 
and therefore surface brightness, to model predictions based on $f(N_{{\rm H  I}},X)$ 
and the KS relation. 
It does not, however, work in the present context of positive detections over a range of surface
brightnesses. For this we require a differential expression for {\rhodot}, rather than a cumulative 
version used by \citet{Wolfe:2006p474}.
Specifically, assuming that LBGs are at 
the center of DLAs, we wish to predict the rest-frame FUV emission of DLAs for
a range of efficiencies of star formation in such a way that we can distinguish 
between possible different efficiencies for each surface brightness interval. 
We find that $d${\rhodot}$/dN$ accomplishes
this by yielding unique non-overlapping predictions for each efficiency.
Each differential interval of {\rhodot} represents a ring around the LBGs corresponding to 
a surface brightness and a solid angle interval subtended by each ring. This surface 
brightness corresponds to the column density of gas corresponding to some radius in the 
radial surface brightness profile and is responsible for the emission
covering that area on the sky. If this gas is neutral atomic-dominated \ion{H}{1} gas,
then we can predict the expected $d${\rhodot}$/dN$ using $f(N_{{\rm H  I}},X)$
and the KS relation.

Specifically, to obtain $d${\rhodot}$/dN$, we differentiate Equation (\ref{eq:rhostardotN})
with respect to $N$. To do so, we need $g(N_{\perp},X)$, since
Equation (\ref{eq:rhostardotN}) depends on Equation (\ref{eq:JofN}),
which depends on $g(N_{\perp},X)$. We find $g(N_{\perp},X)$ from
$f(N_{{\rm H  I}},X)$ to
obtain a general form of the equivalent double power-law fit for 
$g(N_{\perp},X)$ when $N_\perp < N_0$ using Equation (\ref{eq:fNNperp}). 
We find that

 \begin{equation}
g_{{H \rm I}}(N_\perp,X)=k_{3}(\alpha+3)\left(\frac{N_\perp}{N_{0}}\right)^{\alpha} \ ;\;\; N_\perp < N_0  \;,\\
\label{eq:gofN}
\end{equation}

 \noindent where $k_3$ and $\alpha$ are the same as in Equation (\ref{eq:fofN}), and 
 $g(N_{\perp},X)$=0 for $N_\perp \ge N_0$.
 We now differentiate Equation (\ref{eq:rhostardotN}) with respect to $N$ to get
 
\begin{equation}
\frac{d\dot{\rho_*}}{dN}=h(N)\left(\frac{H_{0}}{c}\right)\left(\frac{Kk_3}{N_c^\beta}\right)
\left(\frac{\alpha+3}{\beta + 2 + \alpha}\right)
\left(\frac{N_0^{-\alpha}}{N^2}\right)\;\;\;,
\label{eq:drhostardotdN_gen}
\end{equation}

\noindent where 

\begin{equation}
h(N) = \left\{\begin{array}{ll}
N^{\beta +2+\alpha}-N_{\rm min}^{\beta +2+\alpha}  \ ; N < N_0  \;,\\
N_0^{\beta +2+\alpha}-N_{\rm min}^{\beta +2+\alpha} \ ; N > N_0  \;.\\
\end{array}
\right.
\label{eq:hofN_gen}
\end{equation}

\noindent In the case of $\alpha$=$\alpha_{3}$=$-$2.00$\pm$0.05 for $N\le N_0$ \citep{Prochaska:2009p11257}, 
this reduces to  $g(N_{\perp},X)$=$k_{3}(N_{\perp}/N_{0})^{-2}$
 at $N_\perp < N_0$ and $g(N_{\perp},X)$=0 for $N_\perp > N_0$. Also, Equations \ref{eq:drhostardotdN_gen} and \ref{eq:hofN_gen} reduce to, 
 
 \begin{equation}
\frac{d\dot{\rho_*}}{dN}=h(N)\left(\frac{H_{0}}{c}\right)\left(\frac{Kk_3}{\beta N_c^\beta}\right)
\left(\frac{N_0}{N}\right)^2\;\;\;, \; (\alpha=-2)
\label{eq:drhostardotdN}
\end{equation}

\noindent where 

\begin{equation}
h(N) = \left\{\begin{array}{ll}
N^\beta-N_{\rm min}^\beta  \ ; N < N_0  \;,\\
N_0^\beta-N_{\rm min}^\beta \ ; N > N_0  \;,\\
\end{array}\; (\alpha=-2)\;.
\right.
\label{eq:hofN}
\end{equation}

\noindent We note that the expression for $d\dot{\rho_*}/dN$ is independent of
$\alpha_4$. 

We would like to compare $d\dot{\rho_*}/dN$  to the observations, however, 
we cannot measure $d\dot{\rho_*}/dN$ directly. On the other hand,  we can measure $d\dot{\rho_*}/d\langle I_{\nu_0}^{\rm obs}\rangle$, which is easily derived from $d\dot{\rho_*}/dN$. Specifically, we find 

\begin{equation}
\frac{d\dot{\rho_*}}{d\langle I_{\nu_0}^{\rm obs}\rangle}=\left(\frac{d\dot{\rho_*}}{dN}\right)
\left(\frac{d\Sigma_{\rm SFR}}{dN}\right)^{-1}\left(\frac{d\Sigma_{\rm SFR}}{d\langle I_{\nu_0}^{\rm obs}\rangle}\right)\;.
\label{eq:drhostardotdIdef}
\end{equation}

\noindent Since

\begin{equation}
\frac{d\Sigma_{\rm SFR}}{dN}=
\frac{K\beta N^{\beta-1}}{N_c^\beta}\;,
\label{eq:dNdsigma}
\end{equation}

\noindent and

\begin{equation}
\frac{d\Sigma_{\rm SFR}}{d\langle I_{\nu_0}^{\rm obs}\rangle}=\frac{4\pi(1+z)^3\beta}{C}\;,
\label{eq:dsigmadI}
\end{equation}

\noindent we therefore find that


\begin{equation}
\frac{d\dot{\rho_*}}{d\langle I_{\nu_0}^{\rm obs}\rangle}=
h(N)\left(\frac{H_{0}}{c}\right)
\left(\frac{4\pi k_3(1+z)^3}{C}\right)
\left(\frac{\alpha+3}{\beta+2+\alpha}\right)
\left(\frac{N_0^{-\alpha}}{N^{\beta+1}}\right)\;,
\label{eq:drhostardotdI_gen}
\end{equation}

\noindent which for $\alpha=-2$ reduces to 

\begin{equation}
\frac{d\dot{\rho_*}}{d\langle I_{\nu_0}^{\rm obs}\rangle}=
h(N)\left(\frac{H_{0}}{c}\right)
\left(\frac{4\pi k_3(1+z)^3}{\beta C}\right)
\left(\frac{N_0^2}{N^{\beta+1}}\right)\;,
\label{eq:drhostardotdI}
\end{equation}

\noindent where $h(N)$ is the same as in Equations (\ref{eq:hofN_gen}) and (\ref{eq:hofN}). Since 
$N$ is related to $\langle I_{\nu_0}^{\rm obs}\rangle$ through Equation (\ref{eq:sbtoN})
and the KS relation, we find

\begin{equation}
N=\langle I_{\nu_0}^{\rm obs}\rangle^{\frac{1}{\beta}}
\left(\frac{4\pi(1+z)^3\beta {N_c}^\beta}{C K}\right)^{\frac{1}{\beta}}\;,
\label{eq:Nofi}
\end{equation}

\noindent and therefore $d\dot{\rho_*}/d\langle I_{\nu_0}^{\rm obs}\rangle$ is a unique 
function of $\langle I_{\nu_0}^{\rm obs}\rangle$, and thus surface brightness.

\begin{figure}[t!]
\center{
\includegraphics[scale=0.5, viewport=15 5 490 350,clip]{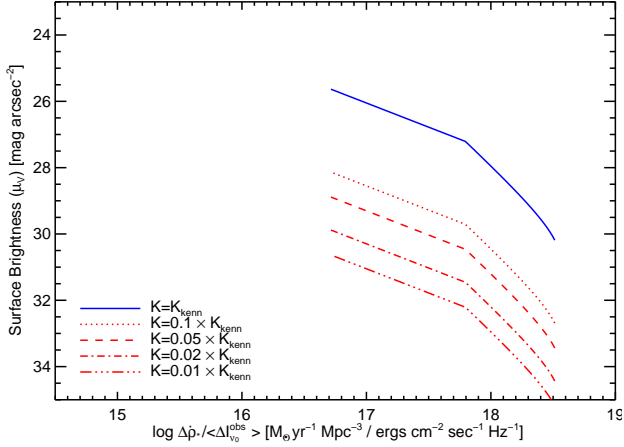}}
\caption{Solid blue curve is the surface brightness ($\mu_V$)  versus the differential comoving SFR density per intensity 
$\Delta \dot{\rho_*}/\Delta\langle I_{\nu_0}^{\rm obs}\rangle$ predicted for 
Kennicutt--Schmidt relation with $K=K_{\rm Kenn}$. The red curves depict what 
$\Delta \dot{\rho_*}/\Delta\langle I_{\nu_0}^{\rm obs}\rangle$ looks like for different values of the normalization
constant $K$, where $K_{\rm Kenn}$=(2.5$\pm$0.5){$\times$}10$^{-4}$ {\smpykpc} 
\citep{Kennicutt:1998p3174, Kennicutt:1998p10742}. 
The curves in this figure predict the amount of star formation 
that should be observed around LBGs for different efficiencies of star formation
from Equations (\ref{eq:drhostardotdI}) and (\ref{eq:Nofi}).
The dotted line is for $K=0.1\times K_{\rm Kenn}$, the dashed line is for 
$K=0.05\times K_{\rm Kenn}$, the the dot-dashed line is for 
$K=0.02\times K_{\rm Kenn}$, and the tripple-dot-dashed line is for $K=0.01\times K_{\rm Kenn}$.
The range of surface brightnesses corresponds to $5\times10^{20}$ cm$^{-2} <N<1\times10^{22} $cm$^{-2}$. 
}
   \label{fig:rhodotstar_sb_nolbg}
\end{figure}

The resulting predictions for the surface brightness, $\mu_V$, versus the 
differential comoving SFR density per intensity,
$d\dot{\rho_*}/{d\langle I_{\nu_0}^{\rm obs}\rangle}$,
is depicted by the blue curve in Figure \ref{fig:rhodotstar_sb_nolbg}. 
The red curves in this figure depict 
$d\dot{\rho_*}/{d\langle I_{\nu_0}^{\rm obs}\rangle}$ for different values of the normalization
constant $K$, where $K_{\rm Kenn}$=(2.5$\pm$0.5){$\times$}10$^{-4}$ {\smpykpc} 
\citep{Kennicutt:1998p3174, Kennicutt:1998p10742}. 
First, we note that $d\dot{\rho_*}/{d\langle I_{\nu_0}^{\rm obs}\rangle}$ for a given $K$
decreases as $\mu_V$ decreases because of the decreasing population of high
column-density DLAs (i.e. DLAs with higher surface brightnesses). 
Second, we note that both {\rhodot} and ${\langle I_{\nu_0}^{\rm obs}\rangle}$ are linearly 
proportional to $K$, and therefore $K$ cancels out in the $x$-direction, 
leaving only the observed $\mu_V$ to vary with $K$ in the $y$-direction. 
We plot $\mu_V$ on the $y$-axis 
to conceptually facilitate the conversion of these results later in the paper
and note that $d\dot{\rho_*}/{d\langle I_{\nu_0}^{\rm obs}\rangle}$ is a function of 
$N$ and therefore $\mu_V$. The curves in this figure predict the amount of star formation 
that should be observed around LBGs for different efficiencies of star formation,
which will be compared to the data in Section 6.2. 
We plot $d\dot{\rho_*}/{d\langle I_{\nu_0}^{\rm obs}\rangle}$
 for the range in $\mu_V$
that corresponds to $N_{\rm min}<N<N_{\rm max}$, 
where $N_{\rm min}= 5\times10^{20} $cm$^{-2}$, and 
$N_{\rm max}= 1\times10^{22}$ cm$^{-2}$. The value of $N_{\rm min}$ is lower than the 
range of threshold column densities of $N_{\perp}^{\rm crit}$ observed for nearby galaxies
\citep{Kennicutt:1998p10742}, 
but at a column density high enough such that  we may start to see star formation occur. 
Also, recent results probing in the outer disks of nearby galaxies observe star formation
at low column densities in atomic-dominated hydrogen gas \citep{Fumagalli:2008p5770, Bigiel:2010b, Bigiel:2010a}.
Regardless, our measurements do not probe column densities down to this level, so
the exact value for $N_{\rm min}$ does not affect the results.

\subsection{Stacking Randomly Inclined Disks}

We wish to compare these theoretical values of  $d\dot{\rho_*}/{d\langle I_{\nu_0}^{\rm obs}\rangle}$,
predicted for DLA gas forming stars according to the KS relation,
to empirical measurements of $\Sigma_{\rm SFR}$ for our LBG sample.  To do this, we require a method to 
convert the radial surface brightness profile in Figure \ref{fig:sb} into $\mu_V$ versus
$d\dot{\rho_*}/{d\langle I_{\nu_0}^{\rm obs}\rangle}$.
For a given ring corresponding to a point in the radial surface brightness profile 
and covering an area $\Delta A$ we can calculate $\Delta \dot{\rho_*}$ from the measured intensity
$\langle I_{\nu_0}^{\rm obs}\rangle$. 
$\Delta \dot{\rho_*}$ is similar to the differential {\rhodot} mentioned above, and is calculated from the flux
measured in an annular ring at some radius from the center of the LBGs.
Specifically, 

\begin{equation}
\Delta \dot{\rho_*} = \frac{\Delta L_\nu N_{\rm LBG}}{C' V_{\rm UDF}}\;,
\label{eq:deltarhodotstar}
\end{equation}

\noindent where $C'$ is the conversion factor from FUV radiation to 
SFR\footnote{$C'$ is in different units than the same factor $C$ in Equation (\ref{eq:sbtoN}).}, 
$C'=8\times10^{27}$ erg s$^{-1}$ Hz$^{-1}$ ($M_\odot$ yr$^{-1}$)$^{-1}$,
 $\Delta L_{\nu}$ is the luminosity per unit frequency interval
 for a ring with area $\Delta A$, 
$N_{\rm LBG}$ is the number of $z\sim3$ LBGs in the UDF, and 
$V_{\rm UDF}$ is the comoving volume of the UDF. As discussed
in Section  3.1, we use a comoving volume of  26436~Mpc$^3$. 
We recognize that $N_{\rm LBG}$ is dependent on the depth of 
images available to make selections, and discuss this completeness issue in Appendix A.3.
The value of $\Delta \dot{\rho_*}$ depends on $\Delta L_{\nu}$, and therefore 
on the inclination angle $i$ for a given measured intensity,
in the case of planes of preferred symmetry.

In order to determine $\Delta \dot{\rho_*}$, 
we average over all inclination angles in our determination of $\Delta L_{\nu}$,
which depends on $\Sigma_\nu^\perp$ described in Section 5.2. 
Specifically, we use 
Equation (\ref{eq:Sigmanu}) in conjunction with 
$\Delta L_{\nu} = \Delta A_\perp\Sigma_\nu^\perp$ to find  $\Delta \dot{\rho_*}$, 
where $\Delta A_\perp$ is the area parallel to the plane of the disk.
First, we rewrite  $\Delta A_\perp$ in terms of $\Delta A$, the projection of $\Delta A_\perp$ perpendicular to the line of sight, 
and find that averaging over all inclination angles yields $\Delta A_\perp = 2 \Delta A$. 
We can then rewrite $\Delta A_\perp$ in terms of $\Delta \Omega$, the solid angle subtended by  
one of the rings from the surface brightness profile,
and $\Delta A$, yielding
$\Delta A_\perp = 2\Delta \Omega d_A^2$. Using this relation, we find

\begin{equation}
\Delta \dot{\rho_*}=\frac{8\pi(1+z)^3{d_A}^2N_{\rm LBG}\Delta F_{\nu_0}^{\rm obs}(r)}{\ln(R/H) C' V_{\rm UDF}}\;,
\label{eq:deltarhodotstarfull}
\end{equation}

\noindent where $\Delta F_{\nu_0}^{\rm obs}(r)$ is the 
observed integrated flux in a ring as a function of the radius $r$.
We then calculate the change in intensity from one ring to the next, $\Delta\langle I_{\nu_0}^{\rm obs}\rangle$, 
to get  $\Delta \dot{\rho_*}/{\Delta \langle I_{\nu_0}^{\rm obs}\rangle}$
by measuring the intensity change across each ring 
by taking the difference between values of the intensity on either side of each point and dividing by two.
In the case that $\Delta\langle I_{\nu_0}^{\rm obs}\rangle$ as calculated above is negative, we take the change in intensity over 
a larger interval.

\begin{figure}
\center{
\includegraphics[scale=0.5, viewport=15 5 495 350,clip]{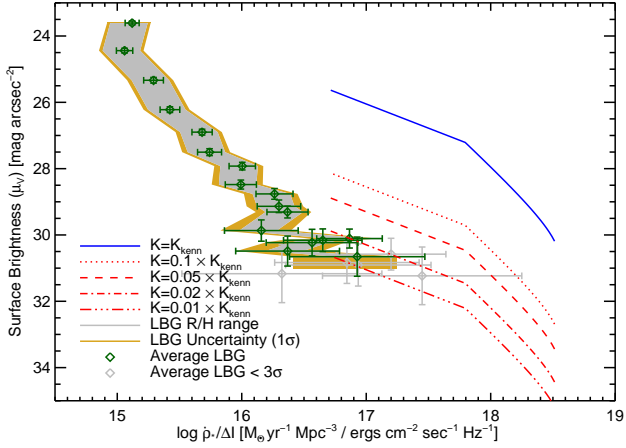}}
\caption{Surface brightness ($\mu_V$) versus the differential comoving SFR density per intensity 
($\Delta \dot{\rho_*}/\Delta\langle I_{\nu_0}^{\rm obs}\rangle$),  comparing the measured emission in the
outskirts of LBGs to the predicted levels for different SFR efficiencies.  
The blue and red lines are the predictions from Figure \ref{fig:rhodotstar_sb_nolbg}. 
The filled gray region represents observed emission in the outskirts of LBGs for a range in aspect ratios, with $R/H$ values ranging from 10 to 100,
and the filled gold region is its 1$\sigma$ uncertainty.
The green diamonds are the average value of the possible aspect ratios,
with the error bars reflecting the measurement uncertainties (including the uncertainty due to the variance in the image composite).
}
\label{fig:rhodotstar_sb}
\end{figure}

Figure \ref{fig:rhodotstar_sb} shows a comparison of the theoretical model from Section 5.2.2 and
the measured values from the radial surface brightness profile. 
The measurements
are for a range in aspect ratios, with $R/H$ values from 10 to 100, 
similar to Section 5.2. We display the data in two complementary ways. First, the 
green diamonds depict results assuming the average value of the possible
aspect ratios,
with the error bars reflecting the measurement uncertainties (including the uncertainty due to the variance in the image composite)
in order to portray the precision of our measurements.
Second, we show a filled region, where the gray represent results for the full range in aspect ratios
and the gold represents the 1$\sigma$ uncertainties on top of that range. This shows
the region that is acceptable for each of those points.  
In both portrayals, we only include uncertainties of the aspect ratios, the variance due to 
stacking different LBGs,  and the measurement uncertainties,  and do not include uncertainties in the FUV light to SFR 
conversion factor ($C$) from Equation (\ref{eq:sbtoN}), or any other such systematic uncertainties.
We truncate the data at a radius of $\sim0.8$ arcsec ($\sim6$ kpc)
corresponding to a $3\sigma$ cut to include only measurements with high signal to noise.
We note that in calculating the S/N values, we include the uncertainties due to the variance of objects in the composite stack.
The data beyond a radius of $\sim0.8$ arcsec are plotted in gray, and while they yield similar results, they are not included 
due to their low S/N.

The results shown in Figure \ref{fig:rhodotstar_sb} do not yet include completeness corrections, 
which we describe in Appendix A.3. 
We present the completeness corrected comparison between the theoretical models of
 $d\dot{\rho_*}/{d\langle I_{\nu_0}^{\rm obs}\rangle}$ for DLAs
with measured values of  $\Delta \dot{\rho_*}/{\Delta \langle I_{\nu_0}^{\rm obs}\rangle}$ 
in Figure \ref{fig:rhodotstar_sb_comp}. 
This shows that, under the hypothesis that the observed extended FUV emission comes from in situ star formation of DLA gas,
the DLAs have an SFR efficiency at $z\sim3$
significantly lower than that of local galaxies\footnote{
We remind the reader that if the working hypothesis is not correct, 
then the results of \citet{Wolfe:2006p474} already constrain the SFR efficiency.}.
In fact, the Kennicutt parameter $K$
needs to be reduced by a factor of 10--50 below the local value, $K=K_{\rm Kenn}$. 
The values of $\Delta \dot{\rho_*}/{\Delta \langle I_{\nu_0}^{\rm obs}\rangle}$  that intersect the predictions 
of the theoretical models for $N=5\times10^{20}$--$1\times10^{22}$ are black crosses and 
have S/N values ranging from $\sim$17 to $\sim$3
suggesting that the measurements are robust. These points correspond to 
radii of $\sim0.4$--$0.8$ arcsec ($\sim3$--$6$ kpc). 
The point with the largest value of $\Delta \dot{\rho_*}/{\Delta \langle I_{\nu_0}^{\rm obs}\rangle}$ of $\sim17.4$
in Figure \ref{fig:rhodotstar_sb_comp} seems to deviate from what otherwise would be a clear trend.
This point corresponds to the point at a radius of $0.72$ arcsec in Figure \ref{fig:sb}, which also differs
slightly from the general decreasing trend in $\mu_V$. However, it is consistent 
within their uncertainties for the surface brightness profile, and we are not concerned about it. 
All the points at radii larger than $\sim0.8$ arcsec also intersect the theoretical models with similar efficiencies, but are not included
as they have lower S/N. The values of $K$ vary for each data point and are not constant
for a given $\mu_V$.
These tantalizing results will be discussed in Section 7, 
and we consider the effects of stacking different samples of LBGs in Appendix B. 

\begin{figure}
\center{
\includegraphics[scale=0.5, viewport=15 5 495 350,clip]{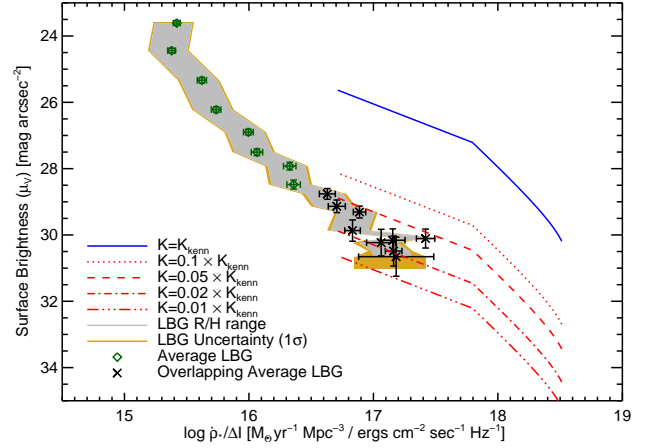}}
\caption{Same as Figure \ref{fig:rhodotstar_sb}, but corrected for completeness. The points that 
overlap with the theoretical models are now black crosses, to emphasize the points that will be used for the rest of this work. 
These points correspond to radii of  0.{\arcsec}405 through 0.{\arcsec}765. 
We omit the low S/N ($<3\sigma$) gray data points from Figure \ref{fig:rhodotstar_sb}. 
By comparing the measurements in this plot to the predictions,
we get the SFR efficiency for each surface brightness.
}
\label{fig:rhodotstar_sb_comp}
\end{figure}

 \subsubsection{Variations in the KS Relation Slope $\beta$}

The SFR efficiencies can also be decreased by lowering the slope $\beta$ of the 
KS relation, while keeping $K=K_{\rm Kenn}$. The value of $\beta$ in the literature ranges from $\beta \sim1.0$
\citep{Bigiel:2008p10261} to $\beta \sim 1.7$ \citep{Bouche:2007p4658}.
Increasing the value of $\beta$ would increase the SFR efficiency,
so we do not consider that here. On the other hand, lower values of $\beta$ would decrease the SFR efficiency
and there are physical motivations to consider
$\beta$ values as low as 1.0 \citep[e.g.,][]{Elmegreen:2002p14454, Kravtsov:2003p11341}.
However, even reducing $\beta$ to values as low as 0.6 does not reduce the SFR efficiency enough 
to match our observations, and there are no physical motivations nor data to justify values of 
$\beta$ lower than 0.6. Lastly, decreasing the value of $\beta$ not only decreases the 
SFR efficiency, but it also decreases the value of $d\dot{\rho_*}/{d\langle I_{\nu_0}^{\rm obs}\rangle}$ 
(see Equation (\ref{eq:drhostardotdI})). While decreasing $\beta$ would move the
blue curve in Figure \ref{fig:rhodotstar_sb_comp} down, it also moves it to the right 
and does not help the models match our observations. 
Therefore, we focus on other mechanisms for reducing the SFR efficiencies by varying 
the value of the parameter $K$.

\subsection{The Kennicutt Schmidt Relation for Atomic-dominated Gas at High Redshift}

As a tool to understand the low SFR efficiencies, and in order to compare our results to
those of \citet{Wolfe:2006p474} and \citet{Bigiel:2010b}, and 
simulations such as  those by \citet{Gnedin:2010p12943}, we 
translate the result from Section 6.2 and Figure \ref{fig:rhodotstar_sb_comp}
to a common set of parameters to obtain a plot
of $\Sigma_{\rm SFR}$ versus $\Sigma_{\rm gas}$. We derive 
this conversion for the emission in the outskirts of LBGs in Section 6.3.1,
and for the DLA upper limits from \citet{Wolfe:2006p474} in Section 6.3.2.

\begin{figure*}
\center{
\includegraphics[scale=0.58, viewport=5 5 500 350,clip]{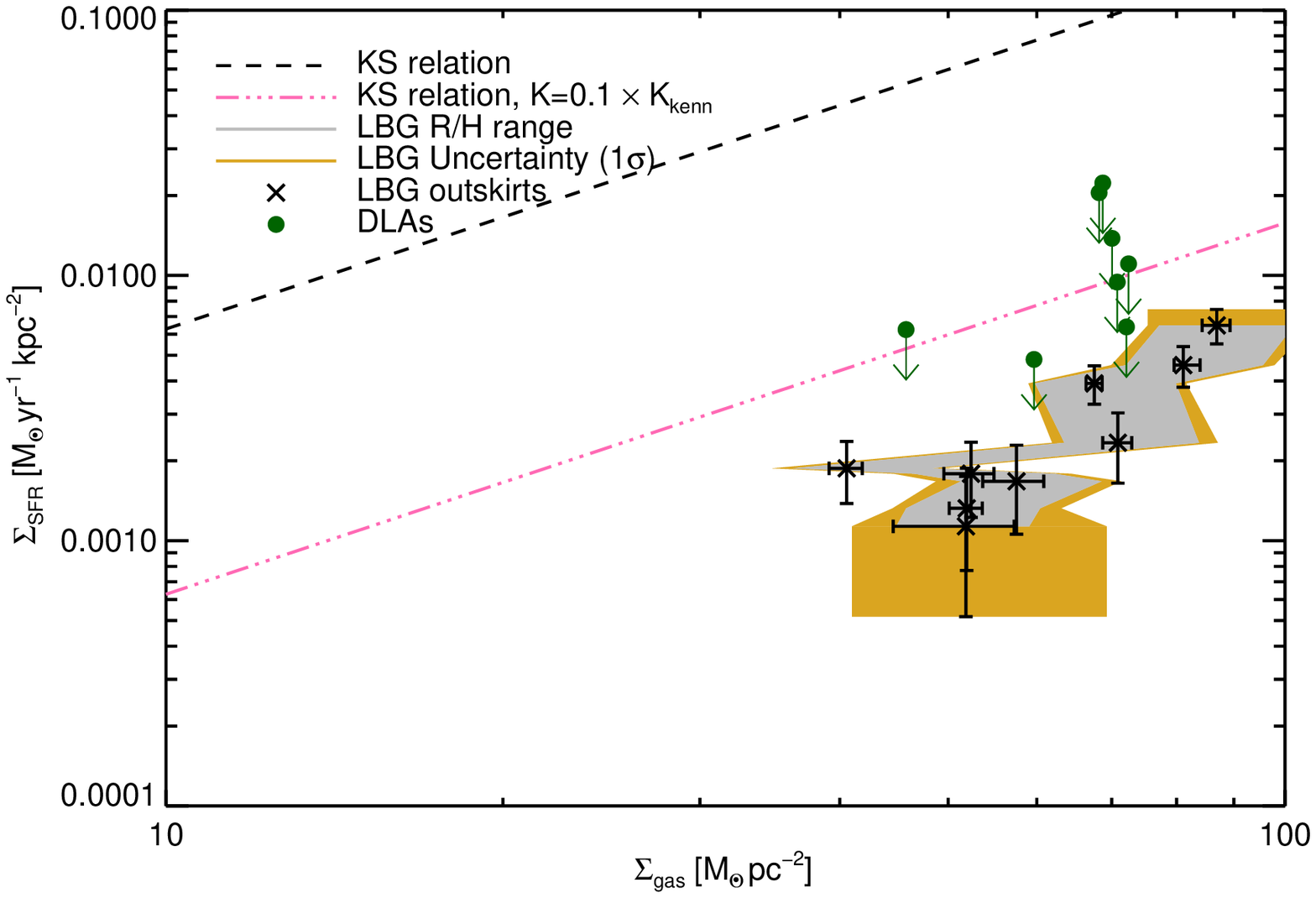}}
\caption{Star formation rate per unit area ($\Sigma_{\rm SFR}$) versus gas density ($\Sigma_{\rm gas}$).
The dashed line represents the 
KS relation with  $K=K_{\rm Kenn}$=(2.5$\pm$0.5){$\times$}10$^{-4}$ {\smpykpc},
while the triple dot-dashed line is for $K=0.1\times K_{\rm Kenn}$.
The gray filled region, the gold filled region, and the black crosses represent the same data as in Figure \ref{fig:rhodotstar_sb_comp}. 
The green data points corresponding to upper limits derived
for DLAs without central bulges of star formation from  \citet{Wolfe:2006p474} converted to work with this plot.
Since DLA sizes are not known, the
upper limits are derived for angular diameters {\thetkern}=4{\arcsec}, 2{\arcsec}, 1{\arcsec}, 
0.{\arcsec}6, 0.{\arcsec}5, 0.{\arcsec}4, 0.{\arcsec}3,
and 0.{\arcsec}25, from left to right. 
The data all fall at or below 10\% of the KS relation, showing a lower SFR efficiency than predicted. 
}
\label{fig:SFR_Sigma_DLA}
\end{figure*}

\subsubsection{Converting the LBG Results to $\Sigma_{\rm SFR}$ versus $\Sigma_{\rm gas}$}

We calculate $\Sigma_{\rm SFR}$ directly using Equation (\ref{eq:sbtoN}), where
the average intensity is obtained from the radial surface brightness profile of the composite LBG 
stack in rings as explained in Section 4.3 (Figure \ref{fig:sb}). 
As $\mu_V$ algebraically increases with increasing radius, 
$\Sigma_{\rm SFR}$ will generally decrease with increasing radius. 
We compute $\Sigma_{\rm gas}$ by first inferring the value of $K$
for each data point that intersects the theoretical curves in the 
$\mu_V$ versus $\Delta \dot{\rho_*}/{\Delta \langle I_{\nu_0}^{\rm obs}\rangle}$ plane, 
since each theoretical curve is parameterized by a fixed value of $K$. 
We precisely find the corresponding efficiency by calculating a grid of models with $K$ varying by 0.001.
We then use this value of $K$ and the KS relation (Equation (\ref{eq:Kenrelation})) to calculate $\Sigma_{\rm gas}$ for each value of 
$\Sigma_{\rm SFR}$ inferred from the measured $\mu_V$. 

In short, we directly measure $\Sigma_{\rm SFR}$ through the emitted FUV radiation, 
and then calculate $\Sigma_{\rm gas}$ for the corresponding $K$ that matches the DLA model.
We note that $\Sigma_{\rm SFR}$ is a direct measurement, while $\Sigma_{\rm gas}$ comes from the DLA
model which is based on the column-density distribution function of DLA gas. 
The result is shown in Figure \ref{fig:SFR_Sigma_DLA}, 
which plots $\Sigma_{\rm SFR}$ versus $\Sigma_{\rm gas}$. 
We truncate the plot to include only data that overlaps with observed DLA gas densities, namely
$N\le1\times10^{22}$ cm$^{-2}$, where $N$ is calculated from Equation (\ref{eq:KenrelationN}) using the 
$K$ value determined for that $\mu_V$. The dashed line black represents the 
KS relation with  $K=K_{\rm Kenn}$=(2.5$\pm$0.5){$\times$}10$^{-4}$ {\smpykpc}, while
the pink triple dot-dashed line represents $K=0.1\times K_{\rm Kenn}$.
The gray filled area with the 1$\sigma$ uncertainty and the black points represent the same data 
as in Figure \ref{fig:rhodotstar_sb_comp}. The green upper limits will be described in Section 6.3.2.

The LBG outskirts in Figure \ref{fig:SFR_Sigma_DLA} clearly have lower SFR efficiencies than 
predicted by the KS relation. In addition, they 
appear to follow a power law that
is steeper than the KS relation at low redshift. However, there is a large scatter 
caused by the uncertainty introduced by the sky-subtraction uncertainty
(see Section 4.2), and the sample variance due to stacking different objects (see Section 4.3). 
We are therefore cautious about fitting a power law to this data by itself, and investigate this trend
 further in Section 7.3.1.

\subsubsection{Converting DLA Data to $\Sigma_{\rm SFR}$ versus $\Sigma_{\rm gas}$}

To convert the DLA points from \citet{Wolfe:2006p474}, we use the same idea as that of 
the LBG data, except that in this case we do not have detected star formation, 
and therefore we use upper limits. Also, rather than using the new framework developed above, we
use the formalism developed for DLAs without central bulges of
star formation \citep{Wolfe:2006p474}. 
We start with Figure 7 of \citet{Wolfe:2006p474}, which basically plots the SFR density due to
star formation in neutral atomic-dominated hydrogen gas (i.e., DLAs) with column densities
greater than $N$ ($\dot{\rho_{*}}(>N)$) versus $\mu_V$\footnote{The $x$-axis of the plot is actually
$\Sigma_{\rm SFR}$, but $\Sigma_{\rm SFR}$ corresponds to a value of $\mu_V$, 
which is easier to understand and makes more sense in this context.}. 
Similar to our Figure \ref{fig:rhodotstar_sb_comp}, Figure 7 of \citet{Wolfe:2006p474}
has DLA models with different Kennicutt parameters $K$.
We follow the same technique of using the intersection
of these models with the data to find $K$ values for each $\mu_V$. However, in this case
the values of $\mu_V$ do not correspond to detected surface brightnesses of LBGs, but rather correspond to 
threshold surface brightnesses, i.e., the lowest values of $\langle I_{\nu_0}^{\rm obs}\rangle$ 
that would be measured for a DLA of a given angular diameter.
Therefore, we cannot calculate $\Sigma_{\rm SFR}$ the same way as in Section 6.3.1. 

To determine $\Sigma_{\rm SFR}$, we first calculate an effective minimum column density, $N_{\rm eff}$, 
corresponding to the threshold surface brightness of \citet{Wolfe:2006p474}.
We do this using
the $K$ values from the intersection points, the thresholds $\mu_V$, and Equation \ref{eq:KenrelationN}. 
We then calculate $\Sigma_{\rm SFR}$, the average over all possible column densities above $N_{\rm eff}$, as
\begin{equation}
\Sigma_{\rm SFR} = \langle \Sigma_{\rm SFR}(>N_{\rm eff})\rangle= \frac{c}{H_0}\frac{\dot{\rho_{*}}(>N_{\rm eff})}{{\int_{N_{\rm eff}}^{N_{\rm max}}J(N')dN'}}\;,
\label{eq:meansfr}
\end{equation}

\noindent where $J(N)$ is the integral in Equation (\ref{eq:JofN}). 
We then calculate $\Sigma_{\rm gas}$ using these $\Sigma_{\rm SFR}$, the above $K$ values, and Equation (\ref{eq:Kenrelation}). 
The resulting values are overlaid on the LBG results in Figure \ref{fig:SFR_Sigma_DLA}. We emphasize that the two results
are independent tests. The DLA upper limits put constraints on the KS relation in the case of no central bulge of star formation, while the 
LBG outskirts data put constraints on in situ star formation in DLAs associated with LBGs. Together, these results show
that the SFR efficiency in diffuse atomic-dominated gas at $z\sim3$ is less efficient than predicted by the KS relation for local galaxies. 

 \begin{figure*}
\center{
\includegraphics[width=0.47\textwidth, viewport=5 5 500 350,clip]{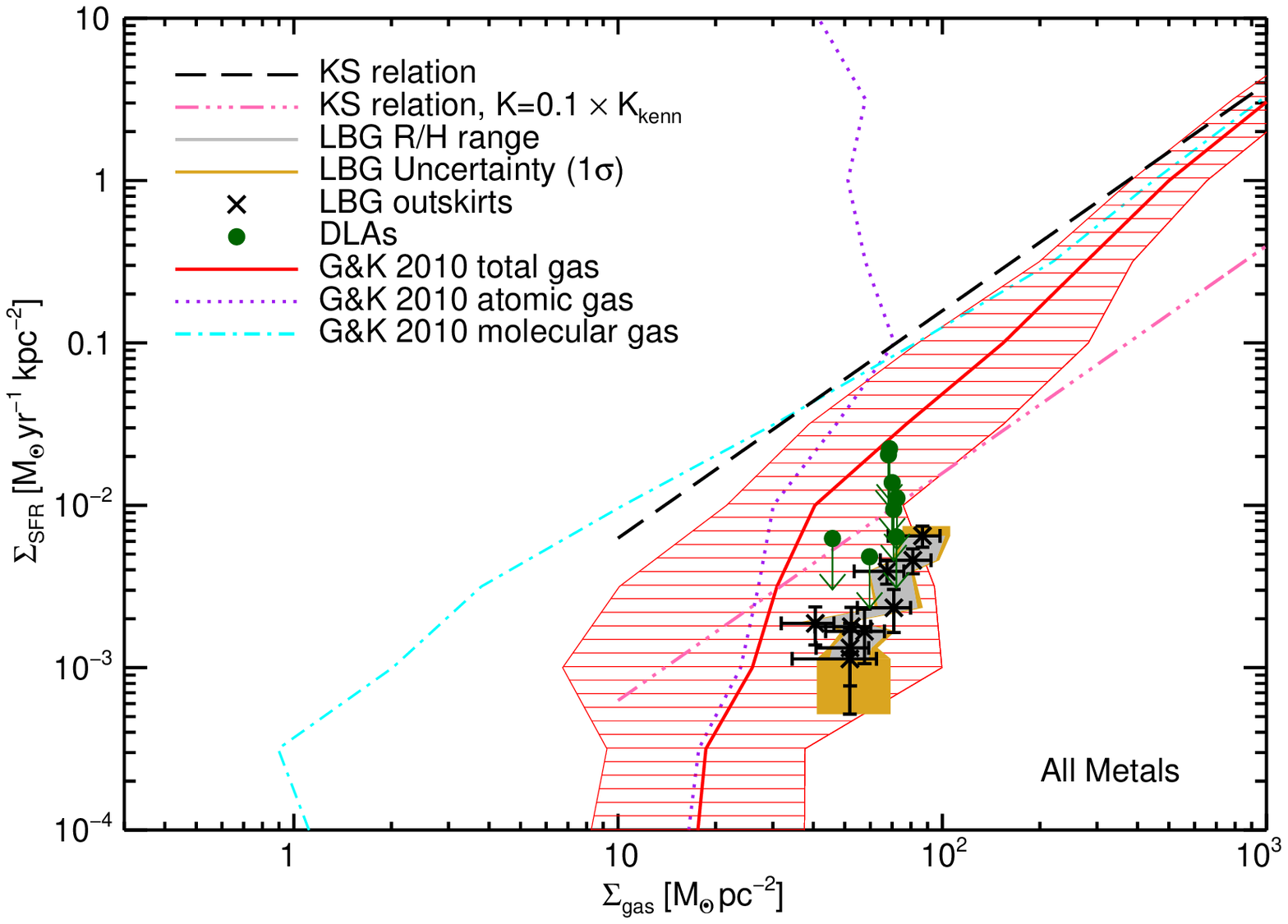}
\hfill
\includegraphics[width=0.47\textwidth, viewport=5 5 500 350,clip]{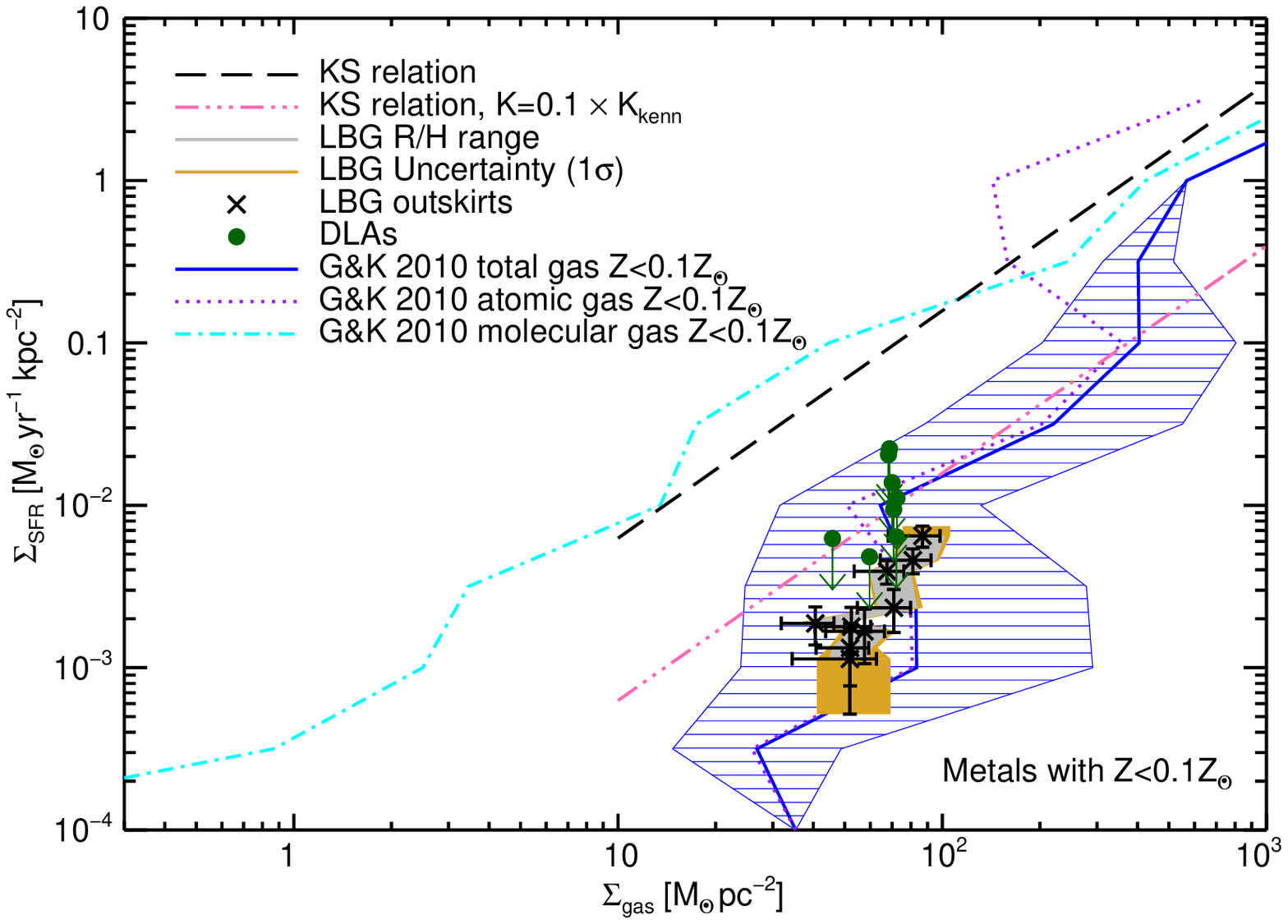}
}
\caption{Star formation rate per unit area ($\Sigma_{\rm SFR}$) versus gas density ($\Sigma_{\rm gas}$) for both the data
as shown in Figure \ref{fig:SFR_Sigma_DLA}, and the KS relation for simulated galaxies at $z=3$ from \citep{Gnedin:2009p13457}. 
The left panel includes gas of all metallicities while the right panel considers only the gas with metallicities below $0.1Z_\odot$.
The solid red and blue lines are the mean relation for the total neutral-gas surface density (atomic and molecular)
and the hatched area shows the rms scatter around the mean. The dotted purple line shows the mean 
KS relation only considering the atomic gas, while the dot-dashed cyan line shows it for the molecular gas. 
The long-dashed line is the Kennicutt relation for local ($z\sim0$) galaxies \citep{Kennicutt:1998p3174}, 
while the triple-dot-dashed line is for $K=0.1\times K_{\rm Kenn}$. 
The gray and gold lines and the black and green points are the data and are the same as in Figure \ref{fig:SFR_Sigma_DLA}.
}
\label{fig:SFR_Sigma_model}
\end{figure*}

\section{Discussion of Results}

We present, for the first time, evidence for LSB 
emission around LBGs on spatial scales
large compared to the LBG cores due to 
in situ star formation in gas associated with LBGs (see Figure \ref{fig:sb}). 
Using the theoretical framework developed in Section 6.1.2, we model this
emission as the average in situ star formation in atomic-dominated gas 
surrounding LBGs  at $z\sim3$. Since most of the atomic-dominated gas at high redshift
is in DLAs, we assume that this is DLA gas. 
We find that under this hypothesis, the efficiency of star formation in the atomic-dominated gas 
is significantly lower than what is expected for predictions
by the KS relation. This is clearly evident in Figure
\ref{fig:SFR_Sigma_DLA}, which compares the local KS relation directly
to the observations. The efficiency of in situ star formation in atomic-dominated gas 
appears to be a factor of $10$--$50$ lower than that of local galaxies that follow KS relation.
 In addition, the SFR efficiency of atomic-dominated gas around galaxies
without bright cores is constrained 
by the thresholds measured by \citet{Wolfe:2006p474}, who impose similar efficiencies
 on the  in situ star formation in DLAs, as shown by the green upper limits in
 Figure \ref{fig:SFR_Sigma_DLA}. Therefore, together with the results from \citet{Wolfe:2006p474},
we constrain the SFR efficiency of {\it all} neutral atomic-dominated hydrogen gas (DLAs) at $z\sim3$.  

There are multiple possible effects contributing to the observed lower SFR efficiencies,
including a higher background radiation field at high redshift, low-metallicity of 
DLAs, and the role of molecular versus atomic hydrogen in star formation. We start the
discussion by comparing our results to the models of low-metallicity 
high-redshift galaxies by \citet{Gnedin:2010p12943} in Section 7.1. We then consider
the roles of molecular and atomic-dominated gas in the KS relation 
in the context of the saturation of atomic dominated in Section 7.2. We investigate
whether there is a variation of the KS relation with redshift in Section 7.3, 
and compare our results at high redshift
with those in the outer disks of local galaxies in Section 7.3.1.
We then consider the effect that a bimodality of the DLA population would
have on the results in Section 7.4, and a caveat on the results based on DLA sizes
in Section 7.5. Lastly, we address the ``Missing Metals'' problem of DLAs in Section 7.6.

\subsection{Models of the Kennicutt Schmidt Relation in High-redshift Galaxies}

Here we discuss two effects contributing to the observed lower efficiencies:
First, at higher redshifts the background radiation field \citep{Haardt:1996p13385} is stronger, 
yielding a higher UV-flux environment. This photodissociates the molecular hydrogen ($\rm H_2$) content of the gas, raising the threshold
for the gas to become molecular, therefore requiring higher gas densities to form stars.
Second, the metallicity of DLAs at high redshift is considerably lower than solar
\citep{Pettini:1994p15931, Pettini:1995p15924, Pettini:1997p15981, Pettini:2002p15950, Prochaska:2003p10084}, 
and therefore has a lower dust content, which is needed to form molecular hydrogen and to shield the gas from
photodissociating radiation. 

Recent theoretical work 
\citep{Krumholz:2008p13733, Gnedin:2009p13457, Krumholz:2009p9796,Krumholz:2009p13722,Gnedin:2010p12943, Gnedin:2011p27997}
suggests that the most important part in determining the amplitude and slope of the KS relation is
the dust abundance, and therefore the metallicity. \citet{Gnedin:2010p12943}  investigate the KS relation at high redshift
using their metallicity dependent model of molecular hydrogen \citep{Gnedin:2009p13457}.
They find that while the higher UV flux does lower the SFR, it also lowers the surface density of
the neutral gas, leaving the KS relation mostly unaffected. However, they find that the lower metallicity,
and therefore lower dust-to-gas ratio, causes a steepening and lower amplitude in the KS relation. This yields
lower molecular gas fractions, which in turn reduces SFR for a given 
gas surface density, $\Sigma_{\rm gas}$. This would yield lower observed SFR efficiencies similar to what is measured in this study.

We compare our results with the $z=3$ model KS relation \citep{Gnedin:2010p12943} in Figure \ref{fig:SFR_Sigma_model},
with a plot showing our data overlaid with the model results, and find that their predictions are consistent with our findings.
This plot is similar to Figure 3 in  \citet{Gnedin:2010p12943}, as we provided them with our preliminary results for comparison.
However, there was previously an error in our implementation of the  theoretical framework, yielding slightly different results than
presented here. Also, here we split the figure into two panels, with the left panel 
including gas of all metallicities for galaxies at $z=3$, and the right panel including only the gas
with metallicities below $0.1Z_\odot$.  Each plot includes the same points as shown in Figure \ref{fig:SFR_Sigma_DLA}.
In addition, the plots show the KS relation for the simulated galaxies at $z=3$ for the total neutral gas (atomic and molecular),
with the solid lines showing the mean value and the hatched area showing the rms scatter around the mean. 
The dotted purple line shows the mean KS relation only considering
the atomic gas, while the dot-dashed cyan line shows it for the molecular gas. 
The dashed black line is the best-fit relation of \citet{Kennicutt:1998p3174} for $z\sim0$ galaxies, 
while the triple dot-dashed line is for $K=0.1\times K_{\rm Kenn}$. 
The blue line in the right panel is closer to the range of observed metallicities observed in DLAs of
 $\sim0.04 Z_\odot$ \citep[][; M. Rafelski et al. 2011 in preparation]{Prochaska:2003p10084}. The use of a metallicity 
 cut of $0.1Z_\odot$ is reasonable, since 
  the mass-weighted and volume-weighted metallicity of the atomic gas 
 is $\sim0.02 Z_\odot$ and $\sim0.03 Z_\odot$ respectively
 (N. Gnedin 2010, private communication),  matching the observed DLA metallicities nicely. 
 In addition, although the dispersion of DLA metallicities is large, the majority of DLAs have metallicities below 0.1 $Z_\odot$, 
 making this a good choice for a cut to compare to DLA gas.

The spatially extended emission around LBGs is consistent with both the total and
low-metallicity gas models in Figure \ref{fig:SFR_Sigma_model}. 
While the uncertainties in both the model and the data are large, taking
the results at face value, we can gain insights into the nature of the gas reservoirs around LBGs. 
The data are a better match to the model with the metallicity cut of $0.1Z_\odot$, coinciding with the mean relation 
for this model.
On the other hand, the emission from gas around LBGs is also consistent within the 1$\sigma$ uncertainties of the 
model including gas with all metallicities for the part of the rms scatter {\it below} the mean. This lower part of the hatched area in 
the left panel of Figure \ref{fig:SFR_Sigma_model} represents the gas at the lower end of the metallicity
distribution shown in \citet{Gnedin:2009p13457}, having a mass-weighted and volume-weighted metallicity of the atomic gas 
 of $\sim0.26$ $Z_\odot$ (N. Gnedin 2010, private communication), which is consistent with that of $z=3$ galaxies having
 an average metallicity of $\sim0.25$ $Z_\odot$ \citep{Shapley:2003p4902, Mannucci:2009p14422}.
Since the observations fall below the mean model for the higher metallicity model, and coincide with the model
for the lower metallicity model, we conclude that the metallicity of the gas is most likely around the 
mean metallicity of the low-metallicity model, $\sim0.04 Z_\odot$, and is definitely below $\sim0.26$ $Z_\odot$, 
the metallicity of the model including gas of all metallicities 
The results therefore imply that the gas in the outskirts of LBGs has lower metallicities 
than the LBG cores. In fact,  the observed SFRs and the implied metallicities from the models 
are fully consistent with the outer regions of LBGs consisting of DLA gas.

\subsection{The Roles of Molecular and Atomic-dominated Gas in the Kennicutt Schmidt Relation}

The reduction in SFR efficiency will also be affected by the competing roles
played by atomic and molecular gas in the KS relation.
There has been some debate
as to whether the KS relation should include both atomic and molecular gas.
All stars are believed to form from molecular gas, and some workers argue
that $\Sigma_{\rm SFR}$ correlates better with molecular gas
than atomic gas \citep{Wong:2002p13491, Kennicutt:2007p4941, Bigiel:2008p10261}. 
On the other hand, other observations show a clear correlation of the total gas surface density, $\Sigma_{\rm gas}$,
with $\Sigma_{\rm SFR}$
\citep{Kennicutt:1989p13472, Kennicutt:1998p3174, Schuster:2007p13697, Crosthwaite:2007p13701},
and the atomic gas surface density,  $\Sigma_{\rm HI}$, with $\Sigma_{\rm SFR}$ \citep{Bigiel:2010b}.

These differences can be understood in the context of the saturation of atomic-dominated gas 
at high column densities. Above a specified threshold of $\Sigma_{\rm HI}$, 
the atomic gas is converted into molecular gas and no longer
correlates with $\Sigma_{\rm SFR}$. This saturation of atomic hydrogen gas above a threshold surface density is clearly observed 
\citep[see Figure 8 of ][]{Bigiel:2008p10261} for local galaxies at $z=0$, and occurs at surface densities of $\sim10$$M_\odot$ pc$^{-2}$ 
\citep{Wong:2002p13491, Bigiel:2008p10261}.
However, below this threshold surface density,
there is a correlation of $\Sigma_{\rm HI}$ and $\Sigma_{\rm SFR}$. This is most clearly seen in the 
results analyzing the {\it outer disks of nearby galaxies} \citep{Bigiel:2010b}, where 
a clear correlation of $\Sigma_{\rm HI}$ with $\Sigma_{\rm SFR}$ is observed. In fact, they find
that the key regulating quantity for star formation in outer disks is the column density of atomic gas. 
Further evidence is seen in the outskirts of M83, where the distribution of FUV flux again follows
the $\Sigma_{\rm HI}$ \citep{Bigiel:2010a}. In fact, \citet{Bigiel:2010a} find that in the outskirts of M83, 
massive star formation proceeds almost everywhere \ion{H}{1} is observed.
While these outer disks must contain some molecular gas in order for star formation to occur,
they are nonetheless dominated by atomic gas with surface densities lower than the saturation threshold seen 
in \citet{Bigiel:2008p10261}. 

The saturation of atomic gas above a threshold surface density is investigated in theoretical models
\citep{Krumholz:2009p13722, Gnedin:2010p12943, Gnedin:2011p27997}, which reproduce the saturation threshold for
atomic gas in local galaxies.
Moreover, the authors find that the threshold surface density for saturation varies with metallicity, where lower metallicity systems
have higher thresholds \citep[see Figure 4 of ][]{Krumholz:2009p13722}. 
In addition, the simulations of $z=3$ galaxies by \citet{Gnedin:2010p12943} show an increased 
saturation surface density of atomic gas of $\sim50$$M_\odot$ pc$^{-2}$, as seen in the purple
dotted line in Figure \ref{fig:SFR_Sigma_model}. The line clearly saturates, with no clear relation 
between $\Sigma_{\rm HI}$ and $\Sigma_{\rm SFR}$ above  $\sim50$$M_\odot$ pc$^{-2}$,
but with a clear relation below this density. 
Since $z=3$ galaxies are typically LBGs
with metallicities of $\sim0.25 Z_\odot$ \citep{Shapley:2003p4902, Mannucci:2009p14422}, 
this is consistent with predictions from \citet{Krumholz:2009p13722}.
The model with only lower metallicity gas (right panel of Figure \ref{fig:SFR_Sigma_model})
saturates at even higher atomic gas surface densities \citep{Gnedin:2010p12943}, 
continuing the relation of the threshold saturation of $\Sigma_{\rm HI}$ with metallicity.
It appears that $\Sigma_{\rm HI}$ tracks $\Sigma_{\rm gas}$ to large values of $\Sigma_{\rm SFR}$,
and no saturation occurs in $\Sigma_{\rm HI}$ until $\sim$ 200--300 $M_\odot$ pc$^{-2}$. 

Comparison of the saturation thresholds found by \citet{Gnedin:2010p12943} with our data in 
the right panel of Figure \ref{fig:SFR_Sigma_model} reveals that the data are below this threshold for saturation
of atomic gas for all values of $\Sigma_{\rm HI}$ for the model with the 0.1$Z_\odot$ cut, 
which most closely matches our results.
We therefore conclude that the measured decrease in SFR efficiency 
is not due to the saturation of atomic-dominated gas. 
In fact, our data confirm that the atomic-dominated gas does not saturate for $\Sigma_{\rm HI}$
of at least $\gtrsim100$$M_\odot$ pc$^{-2}$. 
This number is larger than predicted for LBG metallicities, yet smaller than predicted for DLA metallicities.
This suggests that the average metallicity of the gas in the
outskirts of LBGs is between $0.1Z_\odot$ and $0.25Z_\odot$. This is consistent with the metallicity 
estimate made in Section 5.4 of $ 0.12 Z_\odot$ to $ 0.19 Z_\odot$ based on the metal production rate of 
star formation. 

While molecular gas is very likely needed to form stars, it is also 
clear from results for local outer disks \citep{Bigiel:2010b} and our results 
at high redshift that star formation can be observed even when molecular gas does not 
dominate azimuthal averages in rings of galactic dimensions. 
We do not suggest that  stars are forming from purely atomic gas, but rather that in the presence of high
density atomic gas, there are sufficient molecules to cool the gas such that star formation can be 
initiated by gravitational collapse. 
We predict that more sensitive measurements of molecular gas in local outer disks
 would result in the detection of molecular gas. On the other hand,
recent models by \citet{Low:2010p20450} suggest that molecular hydrogen may not be 
the cause of star formation, but rather a consequence of the star formation. In this scenario,
CO traces dense gas that is already gravitationally unstable and would probably form stars regardless.
More investigations are needed to better understand what is fundamentally required for star formation, and 
our work provides some observational constraints for such work.

We note that even though large amounts of molecular 
gas are not observed in DLAs \citep{Curran:2003p16408, Curran:2004p16277}, 
the molecular gas has a very small covering fraction, and therefore is unlikely to be 
seen along random sight lines \citep{Zwaan:2006p13760}. 
We also note that \citet{Bigiel:2010b} find that the FUV emission reflects the recently formed stars without 
large biases from external extinction. We similarly do not expect much extinction in the outer
parts of the LBGs, as DLAs have low dust-to-gas ratios \citep{pett04, Frank:2010p13890}.
Therefore, unless the gas in the outskirts of LBGs is due to star formation in molecular-dominated gas,
we do not expect large extinction corrections to be necessary. 

\subsection{Is There a Variation of the Kennicutt Schmidt Relation with Redshift?}

Recent studies of star-forming galaxies at high redshift have suggested that the KS relation does not vary with redshift
\citep{Bouche:2007p4658, Tacconi:2010p12798, Daddi:2010p13822,Genzel:2010p19560}. These studies do a
careful job of comparing $\Sigma_{\rm SFR}$ and the molecular gas surface density, $\Sigma_{\rm H_2}$, of
high- and low-redshift systems, and find that the galaxies fit a single KS relation.
This is starkly different than our finding a lower SFR efficiency at high redshift. Furthermore, these studies differ from
simulations of high-redshift galaxies that do find a reduced SFR efficiency \citep{Gnedin:2010p12943}.

These differences may be due to comparisons of different types of gas.
 In our results, we consider atomic-dominated gas as found in DLAs,
 while the other observational studies are focused on molecular-dominated 
 gas with galaxies having high molecular fractions and higher metal abundances.
Similarly, \citet{Bigiel:2010b} also found a lower SFR efficiency in the outskirts of otherwise
normal local galaxies when probing atomic-dominated gas. 
There are two different possible scenarios that explain the results.
First, the KS relation for atomic-dominated gas
follows a different KS relation than the KS relation for molecular-dominated gas.
This possibility would explain the observed lower efficiencies of star formation in
(1) the outskirts of $z\sim3$ LBGs as measured here,
(2) the DLAs without star-forming bulges as measured by \citet{Wolfe:2006p474}, and 
(3) the outskirts of local galaxies as measured by \citet{Bigiel:2010b}. 

On the other hand, the simulations by \citet{Gnedin:2010p12943} focus on galaxies with low metallicities similar
to those observed of LBGs ($\sim0.26$ $Z_\odot$) and find reduced SFR efficiencies
in both their molecular-dominated and their atomic-dominated gas. They also find that the 
efficiency is directly related to the metallicity of the gas, suggesting that the decreased star formation
efficiency is most likely due to decreased metallicities. These models accurately predict the 
SFR efficiencies we measure in the outskirts of the LBGs if they are associated with 
DLAs. The outskirts of the local galaxies measured by \citet{Bigiel:2010b} are also generally of 
lower metallicities \citep[e.g. ][]{GildePaz:2007p14632, Cioni:2009p14585, Bresolin:2009p14654}, 
so the decreased efficiencies could also be due to the lower metallicity.
The SFR efficiencies in \citet{Bigiel:2010b} are similar to this study, and we discuss this below in Section 7.3.1. 

In addition, \citet{Bigiel:2010a} find a clear correlation of the FUV light (representing 
star formation) with the location of the \ion{H}{1} gas in M83. In the inner region they find a much steeper radial decline in $\Sigma_{\rm SFR}$
than in $\Sigma_{\rm H_I}$, while in the outer region of the disk, $\Sigma_{\rm SFR}$ declines less steeply, if at all
\citep[see Figure 4 of ][]{Bigiel:2010a}. Similarly, the metallicity in the inner part of M83 drops pretty steeply and then flattens out
at larger radii \citep{Bresolin:2009p14654}. These results are consistent with the scenario that the metallicity is driving the efficiency 
of the SFR.

While it is not yet clear what causes the decrease in SFR efficiency in DLAs, the results
suggest it is either due to a different KS relation for atomic-dominated gas or due to the metallicity of the gas rather than the redshift. 
Given the excellent agreement of our results with the predictions by \citet{Gnedin:2010p12943}, and the suggestive results of the outer
regions of M83, we give extra credence to the metallicities driving the SFR efficiencies. 
In fact, these two effects may be the same, as the outskirts of galaxies generally probably have lower metallicities, 
and therefore lower SFR efficiencies. 
While all the gas at high redshift may not have a reduced efficiency, care must be taken when using the KS relation
in cosmological models, as the properties of gas vary with redshift, thereby affecting the SFR efficiencies.

\begin{figure*}
\center{
\includegraphics[scale=0.65, viewport=15 5 495 350,clip]{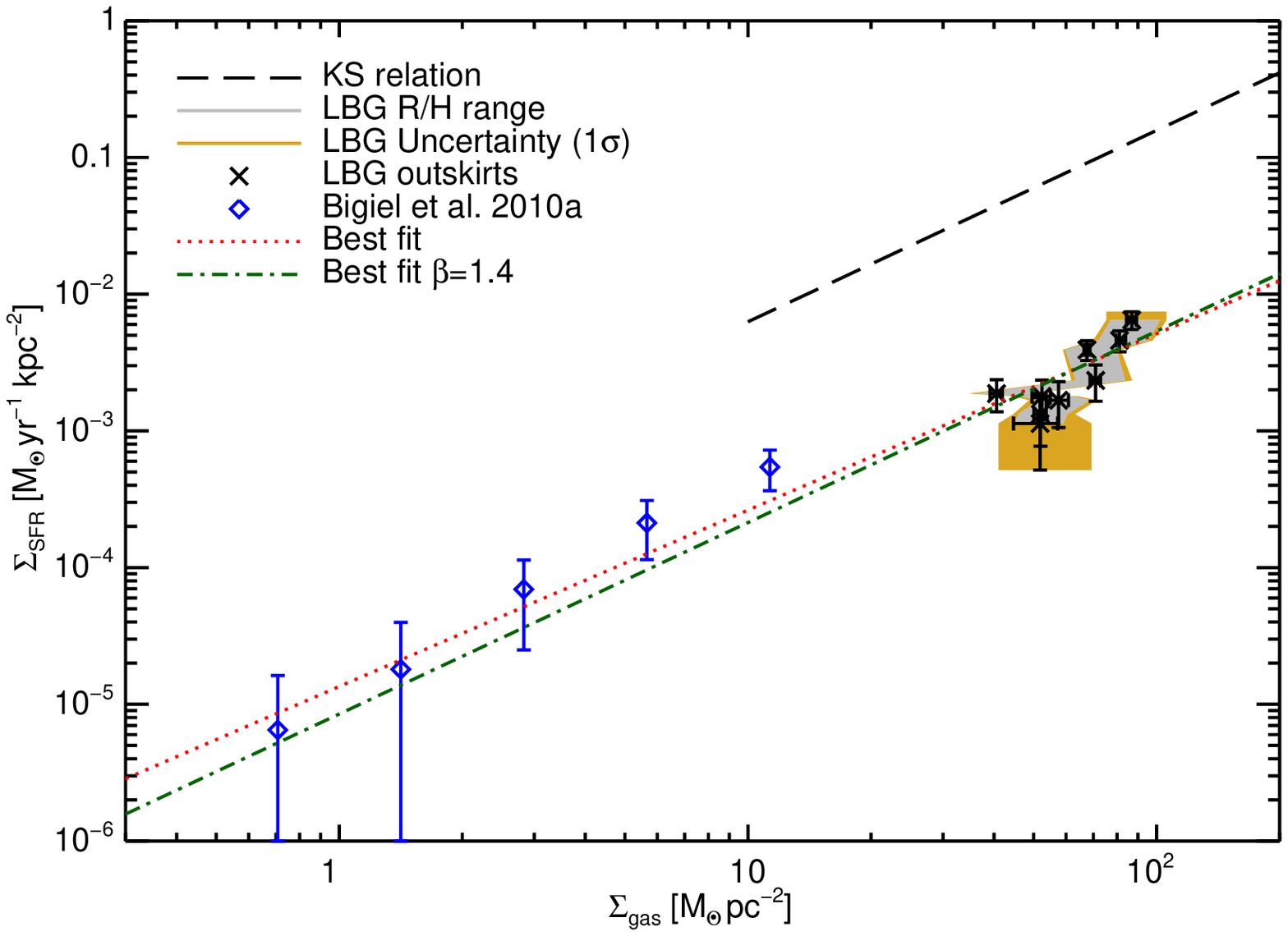}}
\caption{Star formation rate per unit area ($\Sigma_{\rm SFR}$) versus gas density ($\Sigma_{\rm gas}$) for the data
as shown in Figure \ref{fig:SFR_Sigma_DLA} and the results from the outer regions of spiral galaxies from \citep{Bigiel:2010b}.
The blue diamonds represent the best estimate for 
the true relation of $\Sigma_{\rm SFR}$ and $\Sigma_{\rm HI}$ from \citet{Bigiel:2010b} accounting for their sensitivity, and the error bars
represent the scatter. These points have been converted to use the same IMF and FUV--SFR conversion as used in this work.
The long dashed line is the Kennicutt relation for local ($z\sim0$) galaxies \citep{Kennicutt:1998p3174}.
The gray and gold lines and the black points are the data as shown in Figure \ref{fig:SFR_Sigma_DLA}.
The dotted red line represents a power-law fit to the data with both the normalization and the slope as free parameters,
and the green dot-dashed line is a power-law fit with the slope set to $\beta=1.4$, as described in Section 7.3.1. 
}
\label{fig:SFR_Sigma_bigiel}
\end{figure*}

\subsubsection{Comparison of the $z\sim3$ SFR Efficiency with Local $z=0$ Outer Disks}

The environment for star formation at large galactic radii is significantly different than in the inner regions of star-forming galaxies, 
having lower metallicities and dust abundances, consisting of more \ion{H}{1} than H$_2$ gas, and being spread out over large volumes. 
These environmental factors undoubtedly have effects on the conversion of gas into stars, and perhaps the SFR efficiency. 
Similar to our results at high redshift, the SFR efficiency in the local ($z=0$) outer disks is also less efficient than the KS relation \citep{Bigiel:2010b}.
Figure \ref{fig:SFR_Sigma_bigiel} plots our results of \ion{H}{1} dominated gas in the outskirts of LBGs at $z\sim3$ from Figure \ref{fig:SFR_Sigma_DLA}, 
in conjunction with
the measurements of \citet{Bigiel:2010b} of the outskirts of local spiral galaxies. These measurements combine data from the 
\ion{H}{1} Nearby Galaxy Survey and the {\it GALEX} Nearby Galaxy Survey to measure both the atomic hydrogen gas surface density and the 
FUV emission tracing the SFR in 17 spiral galaxies. 
The blue diamonds in Figure \ref{fig:SFR_Sigma_bigiel} represent the best estimate for 
the true relation of $\Sigma_{\rm SFR}$ and $\Sigma_{\rm HI}$ from \citet{Bigiel:2010b} accounting for their sensitivity, 
and the error bars
represent the scatter in the measurements. The SFRs determined in \citet{Bigiel:2010b} from the FUV luminosity use a
Kroupa-type initial mass function (IMF) and the \citet{Salim:2007p11295} FUV--SFR calibration, while we use a Salpeter IMF and the FUV--SFR calibration from
\citet{Madau:1998p13049} and \citet{Kennicutt:1998p10742}. 
For comparison with our LBG results, we convert the SFRs from \citet{Bigiel:2010b} to Salpeter by multiplying by a factor of 1.59 and to
the \citet{Kennicutt:1998p10742} FUV--SFR calibration by multiplying by a factor of 1.30.

While these populations probe very different redshifts and gas densities, they both probe diffuse atomic-dominated 
gas in the outskirts of galaxies and have low metallicities. 
In fact, the differing redshifts enable a comparison of 
drastically different gas surface densities not possible at low redshift alone, 
due to the saturation of \ion{H}{1} gas above a threshold surface density as described in Section 7.2. 
Specifically, \citet{Bigiel:2010b} cannot probe significantly higher gas densities, as at $z=0$ the \ion{H}{1} gas saturates
at threshold surface densities of $\sim10$$M_\odot$ pc$^{-2}$  \citep[see Figure 8 in][]{Bigiel:2008p10261}. 
On the other hand, 
as described in Section 7.2, at $z\sim3$ the \ion{H}{1} gas does not saturate until much higher gas surface densities. 
However, our measurements at $z\sim3$ are also limited in the other direction, as we cannot probe 
lower gas surface densities due to insufficient surface-brightness sensitivity, and therefore we are left with non-overlapping results for
$10 M_\odot$pc$^2 \lesssim \Sigma_{\rm gas} \lesssim 30 $$M_\odot$pc$^2$.

It is not likely a coincidence that the SFR is less efficient in the outer regions of galaxies at both
low and high redshifts. It may be that the SFR is less efficient in the outer regions of galaxies at all redshifts, 
and this possibility needs to be investigated. 
There is evidence that the metallicities in the inner regions of galaxies observed in emission are higher than that in the outer parts sampled by 
quasar absorption line systems at $z\sim0.6$ \citep{Chen:2005p14658, Chen:2007p14659}.
If the SFR efficiency depends on the metallicity, then we would expect lower efficiencies in the outer parts of these $z\sim0.6$ galaxies as observed.

Comparing the data at low redshift with those at high redshift, we find that they appear to fall on a straight line in log space. 
We therefore fit a power law to the outer disk measurements in Figure \ref{fig:SFR_Sigma_bigiel}, using the KS relation in Equation 
(\ref{eq:Kenrelation}) and setting $\Sigma_{\rm c}=1~$$M_\odot$pc$^{-2}$. Leaving both the normalization and the slope as free parameters, 
our least-squares solution results in a normalization of $K=(1.4\pm0.7)\times 10^{-5}$ {\smpykpc} and a slope of ${\beta}$=1.3$\pm$0.1, 
and is plotted as a dotted red line in Figure \ref{fig:SFR_Sigma_bigiel}.
In addition, we also fit a power law 
where we set the slope to the same value as in the KS relation \citep{Kennicutt:1998p3174, Kennicutt:1998p10742},
namely ${\beta}$=1.4. This fit has a normalization of $K=(8.5\pm0.7)\times 10^{-6}$ {\smpykpc}
and is plotted as a dot-dashed green line in Figure \ref{fig:SFR_Sigma_bigiel}.
The fit is consistent for both the low- and high-redshift galaxy outskirts, and the normalization 
is significantly lower than that of the KS relation, with a $K$ value that is 0.03 times the local value.  
We note that the slope of the LBG outskirts is larger than the slope for the outskirts of local galaxies.
The metallicities of the outskirts of local galaxies are probably somewhat higher than those in the outskirts of LBGs, assuming
they consist of DLA gas, and therefore this difference may be a result of differing metallicities. 

We acknowledge that we may be comparing different environments when considering the high and low redshift outer disks, 
but they may be similar enough to tell us about star formation in the outskirts of galaxies. Indeed, both populations exhibit
similar SFR efficiencies. Measurements of the SFR efficiency across a range of redshifts are needed to further explore if variations of the slope are real 
and evolve over time. 

\subsection{Bimodality of DLAs}

There is evidence to suggest that there are two populations of DLAs (high cool and low cool), since
the distribution of cooling rates of DLAs is bimodal \citep{Wolfe:2008p5160}.  These two populations
have significant differences in their cooling rates (and therefore SFRs), velocity widths, metallicity, 
dust-to-gas ratio, and \ion{Si}{2} equivalent
widths. The gas in the high cool population cannot be heated by background and/or  in situ star formation alone,
and so \citet{Wolfe:2008p5160} suggest that the high cool population is associated with compact
star-forming bulges, or LBGs. These high cool DLAs typically have metallicities ($Z=0.09\pm0.03~Z_\odot$), 
higher than the average for DLAs \citep{Wolfe:2008p5160}.
The low cool population may also be associated with LBGs, 
although it is possible that this population is heated solely by in situ star formation. 
These low cool DLAs generally have lower than average metallicities \citep[$Z=0.02\pm0.01~Z_\odot$;][]{Wolfe:2008p5160}.

Given that our measurements in the outskirts of LBGs are consistent with both metallicities, it is possible 
that all DLAs are in the outskirts of LBGs. Alternatively, it would also be 
logical to associate the high cool population of DLAs with the gas responsible for the 
star formation in the outskirts of the LBGs, and the low cool population
would then naturally be represented by the DLA upper limits from \citet{Wolfe:2006p474}.
While those DLAs are also likely associated with galaxies, their lower surface brightness may result in this being included in the 
\citet{Wolfe:2006p474} sample. 

If we are mainly probing the high cool DLAs with the extended LSB emission around LBGs, and 
are mainly probing the low cool DLAs in the upper limits from \citet{Wolfe:2008p5160}, then the number count
statistics would need to be modified to account for the different percentages of high cool and low cool DLAs. 
However, these percentages are not yet very well constrained. A reasonable first try to modify the DLA models 
is to use the current observational result that about half the DLAs are high cool and half are low cool. 
If only high cool DLAs are included, it would lower the expected {\rhodot}, 
which would increase the efficiencies of the SFR, moving the
data to the left in Figure \ref{fig:SFR_Sigma_model}. Assuming that each of the two DLA populations represents half the DLAs,
we find the data move to the left by about $\sim0.1$ dex. This does slightly worsen the agreement of the outskirts of LBGs with the blue low
metallicity model, but not by much. Also, in this scenario, the gas would have higher metallicities ($Z\sim0.09$), and therefore
we would expect the outskirts of the LBGs to fall to the left of the blue line. 
Regardless, the uncertainties in the models and the data are large enough that they could be compatible with
a wide range of interpretations. In short, our SFR efficiency results do not drastically change
whether or not there is a bimodality in the DLA population.

\subsection{Effects of Different DLA Sizes}

One caveat remains, which is the possibility that {\it if} DLAs are not associated with LBGs,
{\it and} the FWHM linear diameters of DLAs at $z\sim3$ are less than 1.9 kpc, then we have not put
a limit on the  in situ star formation in DLAs. However, as discussed in 
\S3.1 of \citet{Wolfe:2006p474}, all models suggested
so far predict that the bulk of DLAs will have diameters larger than 1.9 kpc
\citep[e.g.,][]{Prochaska:1997p11937, Mo:1998p13300, Haehnelt:2000p13255, Boissier:2003p9660, 
Nagamine:2006p466, Nagamine:2007p462, Tescari:2009p9594, Hong:2010p14985}, 
which is in agreement with the observations \citep{Wolfe:2005p382, Cooke:2010p19462}. 
\citet{Rauch:2008p10873} find spatially extended Ly-$\alpha$ emission from objects with
radii up to 4{\arcsec} ($\sim$30kpc) and mean radii of $\sim1${\arcsec} ($\sim$8 kpc)
which they argue corresponds to the \ion{H}{1} diameters of DLAs. While continuum emission has not been 
detected on these scales, it may have been observed from associated compact objects. In fact,
almost all of the Ly-$\alpha$ selected objects have at least one plausible continuum counterpart
(M. Rauch, 2010, private communication). 

If we assume that the sources of Ly-$\alpha$ photons are distributed throughout the gas,
then we can compare the $\Sigma_{\rm SFR}$ values from \citet{Rauch:2008p10873} to 
ours. In this scenario, the in situ star formation in the DLA gas is the source of  the 
Ly-$\alpha$ photons, which is possibly associated with LBGs detected in the continuum. 
We note that currently there is no evidence for this assumption, and it is counter to the interpretation 
by \citet{Rauch:2008p10873}, who assume that Ly-$\alpha$ radiation originates in a compact 
object embedded in the more extended DLA gas. 
On the other hand, we cannot rule out this possibility and it is intriguing, so we consider it here.
Using the median size of their measured extended Ly$\alpha$
emission of $\sim1${\arcsec}, we convert their SFRs to 
$\Sigma_{\rm SFR}$ of 4$\times$10$^{-4}$ -- 8$\times$10$^{-3}$ $M_\odot$ yr$^{-1}$ kpc$^{-2}$. 
This range in  $\Sigma_{\rm SFR}$ overlaps our measured $\Sigma_{\rm SFR}$ shown
in Figures \ref{fig:sb} and \ref{fig:SFR_Sigma_DLA}, suggesting that we may be measuring 
the same star formation in DLAs in two very different methods. 
If this is the case, then the sizes of DLAs may be even larger than 
otherwise expected. Whether or not this interpretation is correct, 
we are definitely above the minimum size probed by \citet{Wolfe:2006p474} of 1.9 kpc, 
implying that the above mentioned
caveat is unimportant. 
We are therefore confident that we 
have shown in Section 5 and Section 6 
that the SFR efficiency in diffuse atomic-dominated gas at $z\sim3$ is less efficient than for 
local galaxies.

\subsection{The ``Missing Metals'' Problem}

The metal production by LBGs can be compared to the metals observed in 
DLAs. Previously, the metal content produced in LBGs was found to be significantly larger
than that observed in DLAs by a factor of 10 and was called the ``Missing Metals'' problem for DLAs
\citep{Pettini:1999p14867, pett04, Pettini:2006p11353, 
Pagel:2002p14992, Wolfe:2003p2460, Bouche:2005p14988}. 
However, \citet{Wolfe:2006p474} found that when taking into account a reduced 
efficiency of star formation as found in DLAs at $z\sim3$, then the metal 
over production by a factor of 10 changed to one of underproduction by a factor of three.

In this study, we calculated the metal production in the outskirts of LBGs
and found that the metals produced there yield metallicities 
in the range of $0.12\pm 0.05 Z_\odot$ to $0.19\pm0.07 Z_\odot$, depending on
the outer radius for the integration of the SFR (see Section 5.3 and Table \ref{tab2}). 
The average metallicity of DLAs at $z\sim3$ is  $Z\sim0.04 Z_\odot$ \citep[][; M. Rafelski et al. 2011, in preparation]{Prochaska:2003p10084}, 
and the metallicity of the high cool DLAs likely associated with LBGs (see Section 7.4) 
typically has metallicities of $Z=0.09\pm0.03~Z_\odot$. We therefore find that
if all the metallicity enrichment of DLAs were mainly due to in situ star formation 
in the outskirts of LBGs, then there would be no ``Missing Metals'' problem. 

In order for the metallicity enrichment of DLAs to arise primarily from in situ star formation
in the outskirts of LBGs,
we would require some mechanism for the LBG cores, where the metallicity is high, to not enrich the intergalactic medium (IGM) significantly from $z\sim10$ to $z\sim3$ ($\sim 2$ billion years).
However, large scale outflows of several hundred km s$^{-1}$ are observed in LBGs 
\citep[e.g.,][]{Franx:1997p15063,Steidel:2001p4911, Pettini:2002p11290, Adelberger:2003p4643, Shapley:2003p4902, Steidel:2010p19909}. 
Therefore, the lack of enrichment would require that either these outflows do not mix significantly with the circumgalactic medium
(CGM, as defined by \citet{Steidel:2010p19909} to be within 300 kpc of the galaxies) in the given time frame 
or that the outflows do not move sufficient amounts of metal-enriched gas to the outer regions of the LBGs to 
significantly affect the metallicity. We note that we also expect some atomic gas to be in the inner regions of LBGs,
and therefore a subsample of DLA gas may be enriched locally, although with a smaller covering fraction.

While there are many theoretical models and numerical simulations to understand the nature of DLAs
\citep[e.g.,][]{Haehnelt:1998p13270, Gardner:2001p15794, Maller:2001p15828, Razoumov:2006p13906,Nagamine:2007p462, Pontzen:2008p5426, Hong:2010p14985},
none of them are able to match the column densities, metallicity range, kinematic properties, and
DLA cross sections in a cosmological simulation. In addition, other than \citet{Pontzen:2008p5426},
they do not reproduce the distribution of metallicities in DLAs. \citet{Pontzen:2008p5426}
do not yet address the mixing of metals nor the effects of large outflows on these metals, and can therefore not 
address this question either. 
These are essential for understanding the production and mixing of the metals, and
it is therefore unclear at this time if outflows from LBGs mix significantly with the CGM,  or if they move enough metal content from
the inner core to the outskirts to significantly increase the metallicity.
If the outskirts are not significantly enriched by the cores, then
the observed metallicity of DLAs would be consistent with expectations from the in situ star formation in the 
outskirts of LBGs.

\section{Summary and Concluding Remarks}

In this work we aim to unify two pictures of the high redshift universe: 
absorption line systems such as DLAs that provide the fuel for star formation, 
and compact star-forming regions such as LBGs which form the majority of stars. 
Each population provides valuable but  independent information about the early universe
\citep[for reviews, see][]{Giavalisco:2002p4895, Wolfe:2005p382}.
Connecting these two populations helps us to better understand how stars form from gas, 
an important part in understanding galaxy formation and evolution. In doing so, we begin to bridge
two separate but complementary fields in astrophysics.

\citet{Wolfe:2006p474} start to bridge these fields by setting sensitive upper limits on star formation in DLAs 
without compact star-forming regions, finding that the in situ star formation
in DLAs is less than 5\% of what is expected from the KS relation. 
However, they do not constrain DLAs associated with bright star-forming cores such as LBGs.
In the present paper we address this caveat by searching
for spatially-extended star formation in the outskirts of LBGs at $z\sim3$ on scales up to
$\sim$10 kpc. We find the following.\\

\noindent 1. Using the sample of 407 $z\sim3$ LBGs in the UDF from \citet{Rafelski:2009}, 
we create a composite image stack in the $V$ band, corresponding to the rest-frame FUV emission
which is a sensitive measure of the SFR, 
for 48 resolved, compact, symmetric, and isolated LBGs at $z\sim3$ 
(Figure \ref{fig:stack}). We detect spatially extended LSB emission in the outskirts of LBGs, 
as shown in the radial surface brightness profile in Figure \ref{fig:sb}. 
This is evidence for the presence of in situ  star formation in gas in the outskirts of LBGs. \\

\noindent 2. We find that the area covered by DLAs is larger than the area of the outskirts of LBGs for 
SFR efficiencies of $K=K_{\rm Kenn}$ and is consistent with DLAs having SFR efficiencies of $K\sim0.1\times K_{\rm Kenn}$ (Figure \ref{fig:cov_frac}).
On the other hand, the covering fraction of molecular gas is inadequate to 
explain the star formation in the outskirts of LBGs (Figure \ref{fig:cov_frac_h2}). 
This suggests that the outskirts of LBGs consist of atomic-dominated gas, 
supporting the underlying hypothesis of this paper.
In fact, the covering fraction provides evidence that the SFR efficiency of atomic-dominated gas at $z\sim3$ 
is on the order of 10 times lower than for local galaxies. \\

\noindent 3. We integrate the rest-frame FUV emission in the outskirts of LBGs and find that the 
average SFR is $\sim 0.1$ $M_\odot$ yr$^{-1}$ and 
{\rhodot} is $\sim3\times10^{-3}$ $M_\odot$ yr$^{-1}$ Mpc$^{-3}$ (see Table \ref{tab2}). 
Combining our largest possible value of  {\rhodot}  in the outskirts of LBGs
with the upper limit found in \citet{Wolfe:2006p474},
we obtain a conservative upper limit on the total {\rhodot} contributed by atomic-dominated gas of 
{\rhodot}$<$ $9.9\times10^{-3}$ $M_\odot$ yr$^{-1}$ Mpc$^{-3}$.
This corresponds to $\sim10\%$ of the {\rhodot} measured in the inner regions of 
LBGs at $z\sim3$ \citep{Reddy:2008p4837}. \\

\noindent 4. We integrate {\rhodot} in the redshift range
3$\lesssim z \lesssim$10,
and calculate the total metal production in DLAs and get a metallicity of $\sim 0.15 Z_\odot$. This is comparable 
to the metallicity of the high cool DLAs believed to be associated with LBGs with metallicities of $\sim 0.09Z_\odot$. If the large observed outflows
of several hundred km s$^{-1}$ of LBGs do not significantly contaminate the CGM, then the metallicity of DLAs would be 
consistent with expectations from the in situ star formation in the outskirts of LBGs. This is a potential solution to the ``Missing Metals'' problem. \\

\noindent 5. Under the hypothesis that the observed FUV emission in the outskirts of LBGs in from in situ 
star formation in atomic-dominated gas, we develop a theoretical framework connecting the emission observed around LBGs
to the expected emission from DLAs. Such a framework is necessary to interpret the
spatially extended star formation around LBGs. This framework develops a differential expression for 
the comoving SFR density, {\rhodot}, corresponding to a given interval of surface brightness 
using the KS relation and the column-density distribution function (Equation (\ref{eq:drhostardotdI})). 
We also develop a method to convert the measured radial surface brightness profile of the LBG composite into the same 
differential expression for {\rhodot} (Equation (\ref{eq:deltarhodotstarfull})). \\

\noindent 6. We compare the predictions for the surface brightnesses and $d\dot{\rho_*}/{d\langle I_{\nu_0}^{\rm obs}\rangle}$ to the measured
values in Figure \ref{fig:rhodotstar_sb_comp} and find that the two overlap if the efficiency of star formation in neutral atomic-dominated
gas is lower than the local KS relation by factors of 10--50.
Using these reduced efficiencies, we convert our results and those from 
\citet{Wolfe:2006p474} into the standard $\Sigma_{\rm SFR}$ versus $\Sigma_{\rm gas}$ plot generally used to study SFRs (Figure \ref{fig:SFR_Sigma_DLA}). 
We find that $\Sigma_{\rm SFR}$ is lower than the upper limits found by \citet{Wolfe:2006p474} for a similar range in $\Sigma_{\rm gas}$, and both 
results have significantly lower SFR efficiencies than predicted by the local KS relation. \\

\noindent 7. The reduced SFR efficiencies in the outskirts of LBGs
 are consistent with the predictions by \citet{Gnedin:2010p12943} for star formation at $z\sim3$ 
 in neutral atomic-dominated gas with low ($<0.1Z_\odot$) metallicities (right panel, Figure \ref{fig:SFR_Sigma_model}). 
 These models find that the primary cause for the lower SFRs is due to the decreased metallicities, suggesting that this is likely
 the primary driver for the reduced SFR efficiencies. \\
 
 \noindent 8. Our results correspond to gas surface densities below the predicted threshold for saturation of atomic gas at $z=3$ for 
 DLA metallicities, and therefore the low measured SFR efficiencies are not due to the saturation of the atomic-dominated gas. 
 In fact, our data support the theoretical predictions \citep{Krumholz:2009p13722, Gnedin:2010p12943}
  that the atomic-dominated gas does not saturate for $\Sigma_{\rm HI}$ of at least $\gtrsim100$$M_\odot$ pc$^{-2}$. \\

  \noindent 9.  Our finding of star formation in atomic-dominated
 gas in the outskirts of LBGs is similar to the recent results by \citet{Bigiel:2010b} who find that star formation in the outskirts of local ($z=0$) galaxies 
also arises in atomic-dominated gas (Figure \ref{fig:SFR_Sigma_bigiel}). 
In fact, the two results are consistent with the same power law, and both results find that the SFR efficiency in this gas is
lower than expected from the KS relation. 
It is possible that the SFR efficiencies in the outskirts of galaxies may be lower at all redshifts, and this tantalizing possibility should be investigated. \\

\noindent 10. We find that the reduced efficiencies of star formation are likely due to either a different KS relation for atomic-dominated 
gas or due to the metallicity of the gas.
 It is possible that the lower efficiencies are due to atomic-dominated gas following a different KS relation than molecular-dominated gas,
 however, we favor the idea that the metallicity of the gas drives the SFR \citep[e.g.,][]{Krumholz:2009p13722, Gnedin:2010p12943, Gnedin:2011p27997}.
In fact, these two effects may be the same, as the outskirts of galaxies generally probably have lower metallicities, 
and therefore lower SFR efficiencies. 
While the reduced efficiency may not be observed for all gas at high redshift, care must be taken when applying the KS relation, 
as the properties of gas vary with redshift, thereby affecting the SFR efficiencies.  \\

 Moreover, simulations of galaxy formation incorporating star formation need to take these latest results into account, as the local KS relation may not be valid when dealing
 with either atomic-dominated gas or gas with lower metallicities. 
Further observations of star formation in atomic-dominated gas are needed to differentiate between these two possibilities responsible for reducing 
the SFR efficiencies. 
Future studies similar to this one at a range of redshifts will help give a clearer picture, as will further studies of star formation in the outskirts of local galaxies. 
In addition, progress will also hopefully be made by measurements 
of the \ion{C}{2} 158$\mu$m line emission  in DLAs using the Atacama Large Millimeter Array 
\citep[e.g.,][]{Nagamine:2006p466}.

\acknowledgements

The authors thank Frank Bigiel, Jeff Cooke,  Nickolay Gnedin, Andrey Kravtsov, and J. Xavier Prochaska for valuable discussions. 
Support for this work was provided by NSF grant AST 07-09235.
This work was made possible by data taken at the W.M. Keck Observatory, which 
is operated as a scientific partnership among the California Institute 
of Technology, the University of California and the National Aeronautics and 
Space Administration.  The Observatory was made possible by the generous 
financial support of the W. M. Keck Foundation.  The authors recognize and 
acknowledge the very significant cultural role and reverence that the summit 
of Mauna Kea has always had within the indigenous Hawaiian community.  We are 
most fortunate to have the opportunity to conduct observations from this 
mountain.  

{\it Facilities:} 
 \facility{Keck:I (LRIS)}, \facility{HST (ACS, NICMOS)}

\begin{center}
{\bf APPENDIX}
\end{center}

\begin{appendices}

\section{Completeness Corrections}

Throughout the paper, we use the number of $z\sim3$ LBGs in the UDF, $N_{\rm LBG}$. 
However, the number of LBGs detected
depends on the depth of our images, and there are large numbers of LBGs fainter 
than our detection limit that are not identified. These LBGs are part of the 
underlying sample and contribute to
$C_A$, {\rhodot}, and $\Delta \dot{\rho_*}/{\Delta \langle I_{\nu_0}^{\rm obs}\rangle}$.
We describe our completeness corrections for each of these in the following sections. 

\subsection{Covering Fraction Completeness Correction}

The covering fraction of LBGs depends on $N_{\rm LBG}$, 
and we recover the underlying covering fraction for all $z\sim3$ LBGs by applying a
completeness correction which assumes that the undetected LBGs 
have a similar radial surface brightness profile as the brighter detected LBGs. 
We acknowledge that the size of LBGs vary with brightness, but do not have measurements 
of its variation at these faint magnitudes, and therefore make this assumption.
If the fainter LBGs were smaller, then their $\mu_V$ would be larger at smaller radii, 
and it would decrease the slope of the covering fraction completeness correction. This would not 
make much of a difference for the correction in Figure \ref{fig:cov_frac}, 
but it could potentially have a small effect on the completeness correction in 
Figure \ref{fig:cov_frac_h2}.

The first step in calculating the completeness correction is determining
the total number of LBGs expected in each
half-magnitude bin from the number counts derived from the best-fit
Schechter function from \citet{Reddy:2009p6997} in Section 3.1. Since the 
number counts in Figure \ref{fig:numcts} agree well with the number counts
of the sample from \citet{Rafelski:2009}, we are confident that this is the appropriate 
number of expected LBGs. We then subtract the number of detected LBGs in each bin
yielding the number of missed LBGs per half-magnitude bin, $N_{j({\rm mis})}$, where $j$ is 
the index in the sum in Equation (\ref{eq:cfmissed}) representing the half-magnitude bins. 
We repeat this for 20 bins from $V\sim$27 to $V\sim$37 mag. The results are
insensitive to the faint limit, as described below.

We then determine the ratio of the flux that needs a correction 
for each half-magnitude bin, ${F}({\rm mag})_j$,  to
the total flux of the composite LBG, ${F_{comp}}$, 
such that the flux ratio is $R_{j} = {F}({\rm mag})_j/{F_{comp}}$.
For each half-magnitude bin, we scale the profile by $R_{j}$ and then
determine the area, $A_j$, that the scaled profile covers for each $\mu_V$. 
The missed covering fraction, $\left(C_A\right)_{\rm mis}$, as a function of $\mu_V$ is just the 
sum over all half-magnitude bins, namely,

\begin{equation}
\left(C_A\right)_{\rm mis} = \sum^{20}_{j=1}
A_j N_{j({\rm mis})}\;.
\label{eq:cfmissed}
\end{equation}

We treat the completeness correction differently for atomic-dominated 
and molecular dominated gas. 
For the atomic-dominated gas, we only consider the
light from the outskirts of these missed galaxies, and not the inner cores.
On the other hand, for the molecular-dominated gas, we consider light from the entire LBG, 
making no distinction between the outskirts and the inner parts of the LBGs. 
In this scenario, both the cores and the outskirts are composed of molecular gas, so 
the cores also contribute to the surface brightness in the outskirts for each $\mu_V$. 
The gold long-dashed lines in Figures \ref{fig:cov_frac} and \ref{fig:cov_frac_h2} correspond
to the covering fraction for the LBGs corrected for completeness.

We note that while $N_{j({\rm mis})}$ increases 
steeply with increasing magnitude, $R_{j({\rm ratio})}$ 
decreases more steeply, and therefore the completeness correction is dominated
by the bright end. To give an idea of the magnitude range of the LBGs contributing the most to the correction,
we discuss the correction for the entire LBG profile here.
The missed LBGs fainter than $V\sim33$ have basically no effect,
since the surface brightnesses barely overlap the area of interest. We could have truncated
the completeness correction at this magnitude with no changes to our results. 
The amount of correction depends on the $\mu_V$ being considered, 
and half the completeness correction for the faintest $\mu_V$ values
is due to LBGs brighter than $V\sim31$. Additionally, those $V\lesssim31$ LBGs only contribute to 
$\mu_V \gtrsim 28$. In other words, for $\mu_V \lesssim 28$, we could have truncated the 
completeness correction at $V\sim31$ without any change.

Hence, the luminosity function used at the fainter magnitudes is not crucial, 
as those points add negligible amounts to the correction. 
However, if there was a significant deviation in the faint end of the luminosity function
between $27 \lesssim V \lesssim 31$, it would affect our completeness corrections 
when considering both the cores and the outskirts. 
(see Figure \ref{fig:cov_frac_h2}). We note, however,  that even if we only correct for completeness
for missed LBGs with $V\lesssim 29$, where a drastic change in the luminosity function is unlikely,
 an evolution of $f^*$ in $f(N_{\rm H_2})$ of $\sim$40 would be needed 
 to account for the LSB emission in the outskirts of LBGs. 
Therefore, variations in the faint end luminosity function are not responsible for the disagreement 
observed in Figure \ref{fig:cov_frac_h2}.
In the case of the atomic-dominated gas where we only consider the outskirts, the 
completeness correction is even more dominated by the brightest missed LBGs, 
since the fainter outskirts quickly do not overlap the $\mu_V$ of interest, 
making the total correction very small (see Figure \ref{fig:cov_frac}).

\subsection{{\rhodotstar} Completeness Correction}

Our determination of {\rhodotstar} also depends on $N_{\rm LBG}$, and we use 
a similar completeness correction to Appendix A.1 here. We use the same formalism, using
the flux ratio $R_j$ and the number of missed LBGs, $N_{j({\rm mis})}$ as calculated there.
We also once again assume that the undetected LBGs have a similar radial surface 
brightness profile as the brighter detected LBGs.
However, this time in addition to summing over all half-magnitude bins $j$, we also sum
over all $\mu_V$ to get the total {\rhodotstar} missed due to completeness, $\dot{\rho_*}_{\rm mis}$. 
Specifically, we find
 
\begin{equation}
\dot{\rho_*}_{\rm mis} = \sum_{\mu_V}\sum^{20}_{j=1}
\dot{\rho_*}~R_{j}N_{j({\rm mis})}/N_{\rm LBG}\;.
\label{eq:rhodotstarmissed}
\end{equation}

\noindent We note that similar to Appendix A.1, this result is again not sensitive to the correction
at the faint end, making the uncertainty of the luminosity function out there unimportant. 

\subsection{$\Delta \dot{\rho_*}$ Completeness Corrections}

The determination of $\Delta \dot{\rho_*}$ depends on $N_{\rm LBG}$, 
which again depends on the depth of our images.
These missed LBGs are part of the 
underlying sample and contribute to the true value of
$\Delta \dot{\rho_*}$, but are not included in Figure \ref{fig:rhodotstar_sb}. 
We recover the underlying true $\Delta \dot{\rho_*}$ for all $z\sim3$ LBGs by applying a
completeness correction similar to Appendices A.1 and A.2. 
We use the same formalism developed in A.1 and 
once again assume that the undetected LBGs have a similar radial surface 
brightness profile as the brighter detected LBGs.
Just like in Appendix A.1, we determine the flux ratio, $R_j$, and the number of missed LBGs, $N_{j({\rm mis})}$,
where the index $j$ represents half-magnitude bins for different missed LBGs. 
For each bin, we calculate the missed $\Delta \dot{\rho_*}/{\Delta \langle I_{\nu_0}^{\rm obs}\rangle}$ as
a function of $\mu_V$ and then sum over the half-magnitude bins
to get the final correction to be applied to  
 $\Delta \dot{\rho_*}/{\Delta \langle I_{\nu_0}^{\rm obs}\rangle}$. Specifically, we find

\begin{equation}
\left(\frac{\Delta \dot{\rho_*}}{\Delta\langle I_{\nu_0}^{\rm obs}\rangle}\right)_{\rm mis} = \sum^{20}_{j=1}
\frac{\Delta \dot{\rho_*}}{\Delta\langle I_{\nu_0}^{\rm obs}\rangle} R_{j}N_{j({\rm mis})}/N_{\rm LBG}\;.
\label{eq:deltarhodotstardImissed}
\end{equation}

\noindent We note that similar to Appendix B, this result is not sensitive to the correction
at the faint end, making the uncertainty of the luminosity function out there unimportant. 
The completeness corrected $\mu_V$ versus 
 $\Delta \dot{\rho_*}/{\Delta \langle I_{\nu_0}^{\rm obs}\rangle}$  is shown in Figure
\ref{fig:rhodotstar_sb_comp}. 

\section{Comparison of the Subset and Full LBG Stacks}

In Section 3.3, we compared the subset and full LBG samples and found them to have similar 
magnitude, color, and redshift distributions. This implies that the two samples of LBGs have similar SFRs, 
stellar populations, and SFHs, and justifies using only the subset sample in the analysis to avoid
contamination. In addition, the KS relation works the same on galaxies of different 
morphologies and environments, so we do not expect an effect based on sample selection. 
Nonetheless, for completeness we investigate how the results would differ if we had 
stacked the full sample of LBGs rather than the subsample. 
A stack of the full sample of LBGs would include contamination and different morphological
areas as described in Section 3.3, 
and therefore these stacks yield an upper limit to the star formation occurring in the outskirts of LBGs. 

We repeat the analysis for the full sample stack, and find that it yields SFR efficiencies slightly higher
than discussed in Section 6.2 for the subsample stack. Specifically, 
the Kennicutt parameter $K$ needs to be reduced by a factor of 
$\sim 9$ below the local value for the entire outskirt region. The slope of
the surface brightness versus $\Delta \dot{\rho_*}/{\Delta \langle I_{\nu_0}^{\rm obs}\rangle}$
is similar to the model  $d\dot{\rho_*}/{d\langle I_{\nu_0}^{\rm obs}\rangle}$, and therefore
yields an efficiency that does vary with radius. This effect is largely due to contamination
from nearby galaxies, and if we repeat the analysis for half the sample including only 
isolated galaxies, we find a varying SFR efficiency with radius. We conclude that
the correct sample to stack is the subsample stack, but a stack of the full sample
does not drastically increase our SFR efficiencies. Therefore, the result of lower
SFR efficiencies is robust and not due to any sample selections. 

We also repeat the covering fraction analysis of Section 5.2 for the full sample stack and half sample stacks
mentioned above. 
In both cases the SFR efficiencies of atomic-dominated gas
would need to be increased by a factor of about two over the results from the subset sample stack
to be compatible with the covering fraction of DLAs. 
Since these stacks include contamination and morphologically different galaxies, 
molecular-dominated gas from parts of some of these galaxies will be contributing throughout. 
Therefore, these results are also upper limits to the covering fraction of the atomic-dominated
gas. Regardless, even with significant contamination, this analysis also indicates that our results
are robust and not due to sample selections. 
In addition, the molecular-dominated covering fraction based on the full LBG stack 
is still insufficient to account for the emission in the outskirts of the LBGs.

\end{appendices}

\bibliography{refs,byhand}

\end{document}